\documentclass[aip,jcp,groupedaddress,amsmath,amssymb,preprint,longbibliography,floatfix]{revtex4-2}
\usepackage[utf8]{inputenc}
\usepackage[T1]{fontenc}
\usepackage{lmodern}
\usepackage{geometry}
\usepackage{amsmath}
\usepackage{amsfonts}
\usepackage{amssymb}
\usepackage{amsthm}
\usepackage{graphicx}
\usepackage{color}
\usepackage{braket}
\usepackage{yfonts}
\usepackage{bbm}
\usepackage{setspace}
\usepackage{fancyhdr}
\usepackage{mathtools}
\usepackage[textsize=tiny]{todonotes}
\usepackage[dvipsnames]{xcolor}
\usepackage{soul}
\usepackage[colorlinks=true, pdfstartview=FitV, linkcolor=black, citecolor=black, urlcolor=black]{hyperref}
\usepackage{aligned-overset}
\usepackage{comment}
\usepackage{tikz}
\usepackage{bm}
\usepackage{siunitx}
\usepackage{mycommands}

\usepackage{fnpct} %

\graphicspath{{fig}}

\DeclareMathOperator{\sign}{sgn}
\NewDocumentCommand{\sbk}{o m}{%
  \IfNoValueTF{#1}
    {\left[ #2 \right]}%
    {\csname#1\endcsname[#2\csname#1\endcsname]}%
}
\NewDocumentCommand{\bk}{o m}{%
  \IfNoValueTF{#1}
    {\left( #2 \right)}%
    {\csname#1l\endcsname(#2\csname#1r\endcsname)}%
}
\NewDocumentCommand{\cbk}{o m}{%
  \IfNoValueTF{#1}
    {\left\{ #2 \right\}}%
    {\csname#1\endcsname\{#2\csname#1\endcsname\}}%
}
\NewDocumentCommand{\abk}{o m}{%
  \IfNoValueTF{#1}
    {\left< #2 \right>}%
    {\csname#1\endcsname<#2\csname#1\endcsname>}%
}
\RenewDocumentCommand{\bra}{o m}{%
  \IfNoValueTF{#1}
    {\left< #2 \right|}%
    {\csname#1\endcsname<#2\csname#1\endcsname|}%
}
\RenewDocumentCommand{\ket}{o m}{%
  \IfNoValueTF{#1}
    {\left| #2 \right>}%
    {\csname#1\endcsname|#2\csname#1\endcsname>}%
}

\newcommand{\Id}{\hat{\mathbbm{1}}}
\renewcommand{\d}{\mathrm{d}}
\newcommand{\Dp}{\Delta \bm{p}}
\newcommand{\Dq}{\Delta \bm{q}}
\newcommand{\q}{\bm{q}}
\newcommand{\p}{\bm{p}}
\newcommand{\nac}{\bm{d}}
\newcommand{\Bp}{\bar{\bm{p}}}

\newcommand{\Bq}{\bar{\bm{q}}}

\renewcommand{\S}{\bm{S}}

\newcommand{\e}{\mathrm{e}}
\newcommand{\w}{\hat{w}}
\newcommand{\winv}{\hat{w}^{-1}}
\newcommand{\pauli}{\hat{\sigma}}

\usetikzlibrary{decorations.markings}
\usetikzlibrary{decorations.pathmorphing}
\usetikzlibrary{arrows}
\usetikzlibrary{arrows.meta}

\tikzset{
  midarrow/.style={postaction={decorate},
    decoration={markings,
    mark=at position .25 with {\arrow[scale=1.5]{>}},
    mark=at position .75 with {\arrow[scale=1.5]{>}}
    }},
}
\tikzset{
  bwfwarrows/.style={postaction={decorate},
    decoration={markings,
    mark=at position .25 with {\arrow[scale=1.5]{<>}},
    mark=at position .75 with {\arrow[scale=1.5]{<>}}
    }},
}
\tikzset{
  wavy/.style={decorate,
    decoration={snake,amplitude=0.3mm}},
}

\begin{document}

\title{A partially linearized mapping approach to surface hopping (MASH-PLDM) for nonadiabatic dynamics and nonlinear spectroscopy}
\author{Jan Obermeier}
\author{Kasra Asnaashari}
\author{Jeremy O. Richardson}
\email{jeremy.richardson@phys.chem.ethz.ch}
\affiliation{\mbox{Institute of Molecular Physical Science, ETH Zurich, 8093 Zurich, Switzerland}}
\date{\today}

\begin{abstract}
We derive a partially linearized version of the mapping approach to surface hopping (MASH) that provides a new method for simulating nonadiabatic dynamics in molecular systems.
As in the original MASH formalism, trajectories travel on adiabatic surfaces with hops determined by the dynamics of a spin vector on the Bloch sphere.
However, because the new approach is based on the partially linearized density matrix (PLDM),
it has two spin vectors, which may be in the same or different hemispheres of the Bloch sphere.
The former scenario describes an electronic population moving on a single adiabatic surface,
whereas the latter describes a coherence moving on the average surface.
We show that this new method, called MASH-PLDM, leads to improved results over the original MASH dynamics for a variety of scattering problems and spin--boson models, especially in cases where coherences play an important role.
A further benefit of the new methodology is that it can treat multi-time correlation functions such as those required for nonlinear spectroscopy.
We demonstrate this capability with a proof-of-principle simulation of time-resolved pump--probe spectroscopy.
\\
\begin{center}
    \includegraphics[width=0.35\linewidth]{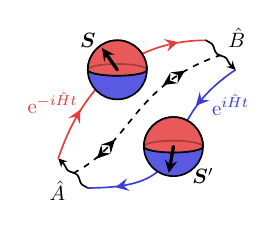}
\end{center}
\end{abstract}
\maketitle

\section{Introduction}

Mixed quantum--classical methods offer a powerful way to simulate nonadiabatic molecular dynamics.\cite{Tully1998MQC,Stock2005nonadiabatic,NonadiabaticBook}
In these approaches, one commonly treats the nuclear degrees of freedom classically, whereas the electronic dynamics are treated using quantum mechanics.
The simplest way to couple quantum and classical dynamics together employs the Ehrenfest approximation, in which the 
electronic state is evolved %
along the nuclear trajectory, which is in turn propagated
according to a mean field defined by the instantaneous electronic state.
The simplicity of this formalism
means that it can be directly generalized 
to simulate %
multi-time correlation functions,\cite{vanderVegte2013nonlinear,Polley2020vibronic,Lieberherr2025nonlinear}
such as those required for nonlinear spectroscopy.\cite{MukamelBook}
However, in practice, the Ehrenfest approximation is known to suffer from a violation of detailed balance,\cite{Parandekar2006Ehrenfest}
making its reliability questionable.

The problems of Ehrenfest theory arise from
the conceptual difficulty in
coupling the Hilbert space of the electrons with the classical phase space of the nuclei in a rigorous way.
For this reason, an alternative and more consistent treatment has been proposed based on mapping the electronic degrees of freedom into classical variables, which together with the nuclear positions and momenta form an extended phase space.\cite{Meyer1979nonadiabatic,Stock1997mapping,Miller2009mapping}
Using these mapping approaches, %
improved methods
can be derived via a \emph{linearization} approximation to the semiclassical initial-value representation (LSCIVR).\cite{Sun1998mapping,Wang1999mapping,Miller2016Faraday,identity,Gao2020mapping}
However, there also exists a related framework called \emph{partial linearization}, in which only the nuclear degrees of freedom are linearized,\cite{Sun1997semiclassical} whereas the forward and backward propagators for the electronic mapping variables are treated in full, resulting in trajectories with one set of nuclear positions and momenta, but two sets of mapping variables.
An example of a method of this class is the partially linearized density matrix (PLDM).\cite{Huo2011densitymatrix,Huo2012MolPhys,Huo2012PLDM}

Whereas we follow the original PLDM derivation using linearized path integrals, similar approaches can be obtained from approximations to the quantum--classical Liouville equation (QCLE).\cite{Kapral1999QCLE,Kapral2015QCL}
This is not surprising as it is known that the QCLE can be derived by linearizing the full path-integral expression.\cite{Shi2004QCLE}
In particular,
PLDM is almost identical to the forward--backward trajectory solution (FBTS) of the QCLE\@.\cite{Hsieh2012FBTS,Hsieh2013FBTS}
In a similar way, it turns out that LSCIVR is strongly related to %
an approximation to
the QCLE called the Poisson-bracket mapping equation (PBME).\cite{Kim2008Liouville}

Benchmarks of PLDM and FBTS found that they were often more accurate than their fully linearized counterparts, LSCIVR and PBME\@.\cite{Huo2011densitymatrix,Huo2012MolPhys,Hsieh2013FBTS}
However, an even more important advantage is that, whereas LSCIVR and PBME can only 
be applied to single-time correlation functions,\footnote{Although LSCIVR is limited to single-time correlation functions, it can be used to simulate nonlinear spectroscopy by explicitly simulating the laser pulses: Gao and Geva, J.\ Chem.\ Theory Comput.\ 16, 6491--6502 (2020).}
partially linearized methods can naturally be extended to multi-time correlation functions.\cite{Provazza2018nonlinear} 

Nonetheless, despite the success of PLDM,
it inherits problems from the underlying mapping approach on which it is based.
It is well known that most mapping approaches have a mean-field character, which, although they tend to improve upon the results of Ehrenfest theory, %
may still not thermalize to the correct equilibrium populations.\cite{thermalization}$^,$\footnote{Although \Refx{Huo2011densitymatrix} seems to imply that PLDM accurately recovers the Boltzmann distribution of the system Hamiltonian in the site basis, exact quantum calculations indicate that these are not in fact the correct equilibrium populations of the full system--bath Hamiltonian\cite{spinPLDM1}}
To some extent, this was improved (although not fully resolved) by replacing the original Meyer--Miller--Stock--Thoss mapping \cite{Meyer1979nonadiabatic,Stock1997mapping,Miller2009mapping} with the spin-mapping formalism\cite{spinmap,multispin} to obtain the spin-PLDM method.\cite{spinPLDM1,spinPLDM2}
This method had particular success in simulating nonlinear spectra such as pump--probe and 2D optical spectroscopy in a set of exciton models.\cite{nonlinear,Ultrafast}
Nonetheless, all these methods are based on dynamics with a mean-field character,
fail to describe wavepacket branching\cite{Huo2012PLDM} %
and can move on unphysical potentials that are formed of linear combinations of multiple states, or even acquire negative populations leading to inverted potentials.\footnote{%
Unlike spin-PLDM and FBTS, the original PLDM method %
avoided the problem of inverted potentials.  However, it did this at the expense of breaking the formal connection to QCLE, as explained in \Refx{Hsieh2012FBTS} and Appendix~\ref{app:QCLE}.}
These problems are not dramatic in simplified harmonic models of condensed-phase systems but could cause disastrous problems in anharmonic molecular systems, particularly when describing bond dissociation.\cite{Bonella2001mapping1}

Within the framework of linearized methods,
recent progress has been made in eliminating the mean-field character of the dynamics.
In particular, the mapping approach to surface hopping (MASH)
introduced a new mapping procedure in which trajectories move on individual adiabatic states with hops initiated deterministically by the dynamics of the electronic mapping variables.\cite{MASH,MASHreview}
The theory is inspired by the success of Tully's fewest-switches surface-hopping (FSSH) method,\cite{Tully1990hopping} but provides a more rigorous foundation based on a similar theory to spin mapping\cite{spinmap} --- the difference being that the electronic populations are defined in terms of discrete rather than continuous variables.
The method shows a number of key advantages over the standard surface-hopping approach, in particular in its description of detailed balance,\cite{thermalization} rates,\cite{MASHrates} rare events, \cite{MASHTPS,MASHTPSDellago} and coherences.\cite{Mannouch2024coherence,MASHcoh}
Additionally, it can be combined with master equations to simulate open quantum--classical systems.\cite{MASHRedfield}
Although the original MASH approach is limited to two-state systems, various multi-state extensions have been proposed\cite{Runeson2023MASH,unSMASH,GranucciMASH}
and applied to simulate the dynamics of excitons and polaritons.\cite{Runeson2024MASH,Runeson2024semiconductors,Runeson2025biology,singletfission,Hasyim2026polariton}
The method is practical and efficient and can thus be applied in conjunction with ab initio electronic-structure methods to simulate the photochemistry of molecules in full dimensionality.\cite{Mannouch2024MASH,cyclobutanone,hutton2024cyclobutatone}

In this work, we will derive a new partially linearized method, %
called MASH-PLDM,
which differs from previous PLDM formulations in the way that the mapping is carried out.
It is based on an ensemble of trajectories
with two spin vectors whose
nuclear force is determined by
both active adiabatic states.
We benchmark the method by calculating single-time correlation functions for a number of model systems,
where the results are shown to improve upon %
the original MASH dynamics.
Finally, we show how the approach can be extended to treat multi-time correlation functions at the heart of nonlinear spectroscopy
and, in this way, simulate a model of a pump--probe experiment.

\section{Theory}

Consider a nonadiabatic molecular system with $f$ nuclear degrees of freedom described by the configuration $\q$ and momentum $\p$.
The Hamiltonian of interest is
$\hat H = \frac{||\hat \p||^2}{2m} + \hat V(\hat \q)$,
where $\hat{V}(\q)$ is the potential-energy operator, also known as the electronic Hamiltonian. %
Note that %
we have scaled the nuclear degrees of freedom such that they all have the same mass, $m$.

At a given nuclear configuration $\q$, the electronic degrees of freedom will be described in a basis of normalized orthogonal adiabatic states, $\ket{\Phi_\mathrm{g}(\q)}$ and $\ket{\Phi_\mathrm{e}(\q)}$,
which diagonalize $\hat{V}(\q)$ with eigenvalues $V_\mathrm{g}(\q)$ and $V_\mathrm{e}(\q)$.
In this way, we have assumed that the relevant dynamics can be truncated to a set of two electronic states, where in the typical scenario, `$\mathrm{g}$' and `$\mathrm{e}$' refer to the ground and excited states.
In this basis, we can write $\hat{V}(\q) = V_0(\q) \Id + V_z(\q) \hat\sigma_z(\q)$,
where $V_0(\q)=\half[V_\mathrm{g}(\q)+V_\mathrm{e}(\q)]$ is the average potential and $V_z(\q)=\half[V_\mathrm{e}(\q)-V_\mathrm{g}(\q)]$ is half the energy gap,
$\Id$ is the identity,
and the (position-dependent) Pauli operators in the adiabatic basis are defined as
\begin{subequations} \label{eq:pauli}
\begin{align}
    \hat \sigma_x(\q) &= \ketbra{\Phi_\mathrm{g}(\q)}{\Phi_\mathrm{e}(\q)} + \ketbra{\Phi_\mathrm{e}(\q)}{\Phi_\mathrm{g}(\q)},\\
    \hat \sigma_y(\q) &=  i \bk[big]{\ketbra{\Phi_\mathrm{g}(\q)}{\Phi_\mathrm{e}(\q)} - \ketbra{\Phi_\mathrm{e}(\q)}{\Phi_\mathrm{g}(\q)}}, \\
    \hat \sigma_z(\q) &= \ketbra{\Phi_\mathrm{e}(\q)}{\Phi_\mathrm{e}(\q)} - \ketbra{\Phi_\mathrm{g}(\q)}{\Phi_\mathrm{g}(\q)}.\label{eq:sigma_z}
\end{align}
\end{subequations}
Note that, in contrast to many studies of electronic spectroscopy,\cite{Shi2008nonlinear,MukamelBook}
these definitions allow for nonadiabatic coupling between the states. %
In particular, due to the position-dependence of the adiabatic states, the Pauli operators have spatial derivatives $\nabla \hat \sigma_x(\q) = 2 \nac(\q) \hat \sigma_z(\q)$, $\nabla \hat \sigma_y(\q) = 0$ and $\nabla \hat \sigma_z(\q) = - 2 \nac(\q) \hat \sigma_x(\q)$,
or more concisely $\nabla\pauli_\mu(\q)=i\nac(\q)[\pauli_y(\q),\pauli_\mu(\q)]$ for $\mu\in\{x,y,z\}$,
where
$\nac(\q) = \abk{\Phi_\mathrm{e}(\q) \middle| \nabla \Phi_\mathrm{g}(\q)}$ 
is the $f$-dimensional nonadiabatic coupling vector.

We wish to compute correlation functions, which are defined in quantum mechanics
as $C_{AB}(t) = \Tr[\hat\rho_\mathrm{n} \hat{A} \, \eu{i\hat{H}t} \hat{B} \, \eu{-i\hat{H}t}]$,
where $\hbar=1$ throughout.
Here, $\hat{\rho}_\mathrm{n}$ is a nuclear density operator and $\hat{A}$ is an electronic operator,
such that the initial conditions are defined by the factorized state $\hat\rho_\mathrm{n}\otimes\hat{A}$.
This is not a limitation, as any coupled nuclear--electronic operator can be expressed as a sum of factorized states in this form.
In the simplest case, $\hat{B}$ is also a pure electronic operator, but the derivation is general enough to also treat pure nuclear operators or mixed nuclear--electronic operators.

\subsection{Partial Linearization} \label{sec:PLDM}

The first part of the derivation is equivalent to previous work on partially linearized methods\cite{Huo2011densitymatrix,Hsieh2012FBTS,spinPLDM1} and is therefore not reproduced here in full.
Instead we simply present the key equations, along with their physical interpretation.
The method is summarized schematically in Fig.~\ref{fig:pldm} and
further details are provided in Appendix~\ref{app:Wigner}.
Note that we use an adiabatic basis, whereas most previous PLDM approaches were derived in the diabatic representation.
In particular, although we use adiabatic states, we retain the kinematic momentum and thus only have to deal with first-order derivative couplings, similarly to Refs.~\onlinecite{Shi2004QCLE} and \onlinecite{Cotton2017mapping}.
This is in contrast to the full adiabatic representation of the Hamiltonian, which uses the canonical momentum and additionally introduces second-order derivative couplings, as described in the Appendix of \Refx{Huo2012PLDM}.

\begin{figure}[h]
    \centering
    \tikzset{
  midarrow/.style={postaction={decorate},
    decoration={markings,
    mark=at position .25 with {\arrow[scale=1.5]{<}},
    mark=at position .75 with {\arrow[scale=1.5]{<}}
    }},
}

\begin{tikzpicture}[>=stealth, ]

\node at (1,-2.5) {$C_{{\color{Blue}A}\green{B}}(t) = \Tr\!\left[{\color{Blue}{\hat{\rho}_\mathrm{n}\hat{A}}\,\color{orange}\e^{i\hat H t}}\,\green{\hat B}\, {\color{purple}\e^{-i\hat H t}} \right]$};

\draw[Blue, thick,wavy,->] (-1,-1) -- (-2,0)
  node[midway,below left] {$\hat{\rho}_\mathrm{n}\otimes\hat{A}$};
  
\draw[purple,thick,midarrow]
  (3,4) .. controls ++(-2,0) and ++(1,3) .. (-2,0)
  node[pos=1, left] {$\bm{S}(0)$}
  node[pos=0.5, above left] {$\bm{S}$}
  node[pos=0, above] {$\bm{S}(t)$};

\draw[orange,thick,midarrow]
  (-1,-1) .. controls ++(4,0) and ++(-3,-2) .. (4,3)
  node[pos=0, below] {$\bm{S}'(0)$}
  node[pos=0.5, below right] {$\bm{S}'$}
  node[pos=1, right] {$\bm{S}'(t)$};
  
\draw[green!50!black,thick,-{stealth[length=3mm]}, wavy] (3,4) -- (4,3)
  node[midway,above right] {$\hat B$};

\draw[black,thick,bwfwarrows]
  (-1.5,-0.5) .. controls ++(2,1) and ++(-3,-1) .. (3.5,3.5)
  node[pos=0, right=0.25cm] {${\Bq}(0)$}
  node[pos=0.5, right] {$\Bq$}
  node[pos=1, left=0.30cm] {$\Bq(t)$}
  ;

\end{tikzpicture}
    \caption{
    The PLDM approximation to the correlation function between $\hat{A}$ at time 0 and $\hat{B}$ at time $t$
    linearizes the nuclear variables of the forward and backward paths into the average path, $\Bq$.
    However, the spin vectors $\S$ and $\S'$, which represent the electronic state, are not linearized and follow different forward and backward paths.
    }
    \label{fig:pldm}
\end{figure}

Discretizing the paths into $N$ timesteps,
the forward and backward propagators are each represented using a path integral with phase-space coordinates $(\q_k,\p_k)$ and $(\q'_k,\p'_k)$ for $k\in\{0,\dots,N\}$. %
We assume that $N$ is chosen large enough %
to keep the discretization error below an arbitrarily low value.
Next, we transform to sum and difference coordinates, $\Bq_k=\half(\q_k+\q'_k)$ and $\Dq_k=\q_k-\q'_k$ (and similarly for the momenta).
Finally, we approximate the action to first order in the difference coordinates, a process known as linearization.
This gives
\begin{align} \label{eq:QCLE}
    C_{AB}(t) \simeq \int \prod_{k=0}^N \d \Bq_k
    \int \prod_{k=0}^N \frac{\d \Bp_k}{(2\pi)^{f}}
    \int \prod_{k=1}^{N} \d \Dq_k 
    \int \prod_{k=0}^{N-1} \frac{\d \Dp_k}{(2\pi)^f}
    \, \rho_\text{n}(\Bq_0, \Bp_0) \tr\sbk{\hat A \hat T' \hat B \hat T} \eu{i \Delta \mathcal{S}_\text{n}},
\end{align}
where we introduce the notation ``$\tr$'' for the trace over the electronic subspace only, whereas ``$\Tr$'' is the full quantum-mechanical trace over both nuclear and electronic spaces.
In addition, the partial Wigner transform (over the nuclei) is defined as
\begin{equation} \label{eq:wigner}
    \rho_\text{n}(\Bq_0, \Bp_0) = \int {\d \Dq_0} \abk{\Bq_0 + \frac{\Dq_0}{2} \middle| \hat \rho_\text{n} \middle| \Bq_0 - \frac{\Dq_0}{2}} \eu{-i \Bp_0 \cdot \Dq_0}.
\end{equation}
In cases where the $\hat{B}$ operator depends on nuclear observables, it is also partially Wigner transformed in an equivalent way and measured at $\hat{B}(\Bq_N,\Bp_N)$.
The phase in \eqn{eq:QCLE} is
\begin{align} \label{eq:DeltaSn}
    \Delta \mathcal{S}_\text{n} = \sum_{k=0}^{N-1} \epsilon \bk{\frac{{\Bq_{k+1} - \Bq_k}}{\epsilon} - \frac{\Bp_k}{m}}\cdot \Dp_k - \sum_{k=1}^{N} \epsilon\bk{\frac{\Bp_{k} - \Bp_{k-1}}{\epsilon} %
    }\cdot \Dq_k,
\end{align}
where the timestep is $\epsilon=t/N$,
and the discretized forward and backward propagators are approximated by
\begin{subequations} \label{eq:hatT}
\begin{align}
    \hat T &= \eu{-\frac{i\epsilon}2 \nabla \hat V(\Bq_{N}) \cdot \Dq_{N}} \, \eu{-i \epsilon \hat V(\Bq_{N})} \cdots \eu{-\frac{i\epsilon}2 \nabla \hat V(\Bq_{1}) \cdot \Dq_{1}} \, \eu{-i \epsilon \hat V(\Bq_{1})}, \label{eq:T}
    \\
    \hat T' &= \eu{+i \epsilon \hat V(\Bq_1)} \, \eu{-\frac{i\epsilon}2 \nabla \hat V(\Bq_{1}) \cdot \Dq_{1}} \cdots  \eu{+i \epsilon \hat V(\Bq_{N})} \, \eu{-\frac{i\epsilon}2 \nabla \hat V(\Bq_{N}) \cdot \Dq_{N}} ,
\end{align}
\end{subequations}
where
\begin{equation} \label{eq:force}
    \nabla \hat V(\q) %
    = \nabla V_0(\q)\Id + \nabla V_z(\q) \hat \sigma_z(\q) - 2 \nac(\q) V_z(\q) \hat \sigma_x(\q).
\end{equation}

The linearization approximation, which we have just applied, becomes exact in the special case of a spin--boson model. \footnote{This is because even though the adiabats themselves are not harmonic, in this case the operator $\nabla^2\hat{V}$ is a constant times the identity in any basis.}
However, in general, it neglects higher-order quantum fluctuations around the path.
This approximate path-integral expression is
equivalent to an explicit solution of the QCLE\@.\cite{Shi2004QCLE}
However, in practice,
carrying out the integrals in Eq.~\eqref{eq:QCLE} numerically 
is not feasible 
in complex molecular systems and thus further approximations are required.

Later, we will integrate over $\Dp_k$ and $\Dq_k$ to obtain equations of motion for the nuclei.
The former results in a delta function which enforces $\Bp_k/m=(\Bq_{k+1}-\Bq_k)/\epsilon$, which %
recovers the intuitive classical relation between the distance moved in a given time and the momentum.
The $\Dq_k$ terms in $\Delta\mathcal{S}_\mathrm{n}$ combined with the $\Dq_k$ terms in the propagators $\hat{T}$ and $\hat{T}'$ will later determine the forces which accelerate the nuclei.
However, we cannot yet integrate over $\Dq_k$ %
as they appear in combination with quantum-mechanical operators.
Therefore, we must first map the operators into an extended phase-space before we can determine the equations of motion.

\subsection{Mapping Approach} \label{sec:mapping}

In both the spin-mapping approach\cite{spinmap} and MASH,\cite{MASH}
the state of a two-level quantum system is mapped onto a Bloch sphere with the (normalized) spin vector $\S=(S_x,S_y,S_z)=(\cos\varphi\sin\theta,\sin\varphi\sin\theta,\cos\theta)$. %
Integrals over the sphere are defined as $\int\d\S\cdots=\frac{1}{2\pi}\int_0^{2\pi}\d\varphi\int_0^\pi\d\theta\sin\theta\cdots$,
which ensures the appropriate normalization (for a two-level system) of $\int \d \S = \tr[\Id] = 2$.

In a similar way to how the Wigner transform maps quantum-mechanical operators onto phase-space functions,\cite{Hillery1984Wigner}
operators of the two-level quantum system, $\hat A$, can be mapped onto functions of the spin vector, $A_{\q}(\S) = \tr\sbk{\hat A \, \hat w_{\q}(\S)}$, using a mapping kernel $\hat{w}_{\q}(\S)$.
Note that because we work in an adiabatic basis, the kernel and hence the mapped functions depend explicitly on the nuclear configuration, $\q$, used to define the basis.
Likewise, phase-space functions can be mapped back to operators using an inverse mapping kernel, $\winv_{\q}(\S)$, with the defining property
$    \hat A = \int \d \S \, A_{\q}(\S) \, \hat w_{\q}^{-1}(\S)$.
Additionally, in order that the identity operator, $\Id$, maps onto unity,
we impose normalization conditions $\tr[\hat w_{\q}(\S)] = 1$ and $\int\rmd\S\,\winv_{\q}(\S)=\Id$.

The fully-linearized spin-mapping method\cite{spinmap} and spin-PLDM\cite{spinPLDM1,spinPLDM2} employed Stratonovich--Weyl kernels,\cite{Stratonovich1957} $\hat{w}_{\q,\mathrm{W}}(\S)=\hat{w}^{-1}_{\q,\mathrm{W}}(\S) = \half\bk{\Id + \sqrt{3}\sum_{\mu\in\{x,y,z\}} S_\mu\hat\sigma_\mu(\q)}$, which are linear in $\S$ and self-dual.
The original MASH method was not defined in terms of kernels at all, but in this work, we have designed the following kernels which reproduce the key behavior of the MASH dynamics:%
\begin{subequations} %
\begin{align}
    \hat w_{\q}(\S) &= \frac12\bk{\Id + 2 S_x \delta(S_z) \hat \sigma_x(\q) + 2 S_y \delta(S_z) \hat \sigma_y(\q) + \sign(S_z) \hat \sigma_z(\q)},\label{eq:MASH_PLDM_kernel}\\
    \hat w_{\q,\text{lin}}^{-1}(\S) &= \frac12\bk{\Id + 2 S_x \hat \sigma_x(\q) + 2 S_y \hat \sigma_y(\q) + 2 S_z \hat \sigma_z(\q)}. \label{eq:MASH_PLDM_inverse}
\end{align}
\end{subequations}
In this way, the Pauli operators are mapped to
$\sigma_x(\S) = 2 \delta(S_{z}) S_{x}$, $\sigma_y(\S) = 2 \delta(S_{z}) S_{y}$, and $\sigma_z(\S) = \sign S_{z}$.
By carrying out the integrals, one can confirm that
the inverse kernel maps them back to the operator form.
We label this particular linear inverse kernel with the subscript ``lin'' as we will later introduce alternative choices.

As we will show, it turns out that the mapping kernel, $\hat{w}_{\q}(\S)$, determines the nuclear forces, whereas the inverse kernel, $\hat{w}_{\q}^{-1}(\S)$, determines the electronic dynamics.
For this reason, the mapping kernel in Eq.~\eqref{eq:MASH_PLDM_kernel} was constructed %
to recover the MASH force function [\eqn{eq:MASHforce}] upon mapping the operator $-\nabla\hat{V}(\q)$ [\eqn{eq:force}].
The inverse mapping [Eq.~\eqref{eq:MASH_PLDM_inverse}] is chosen, like the Stratonovich--Weyl kernels used in spin-PLDM, to treat all three elements of the Pauli operators equivalently and linearly, so as to obtain a unitary spin evolution.
The remaining numerical factors were chosen to ensure that the kernels obey all of the properties listed above. %

We proceed by introducing a notion of time evolution for the spin vector, determined by the transformation properties of the inverse kernel, $\winv_{\q}(\S)$.
It is known that unitary transforms interconvert Pauli matrices among each other like a rotation in three dimensions
[a consequence of the isomorphism between the Lie algebras $\mathfrak{su}(2)$ and $\mathfrak{so}(3)$].\cite{MahlerBook}
Therefore, any unitary transform of the inverse kernel will return another inverse kernel of the same form, but with a rotated spin vector. 
This important relation holds for the inverse kernel (but not the mapping kernel) due to the fact that it is linear in $\S$.

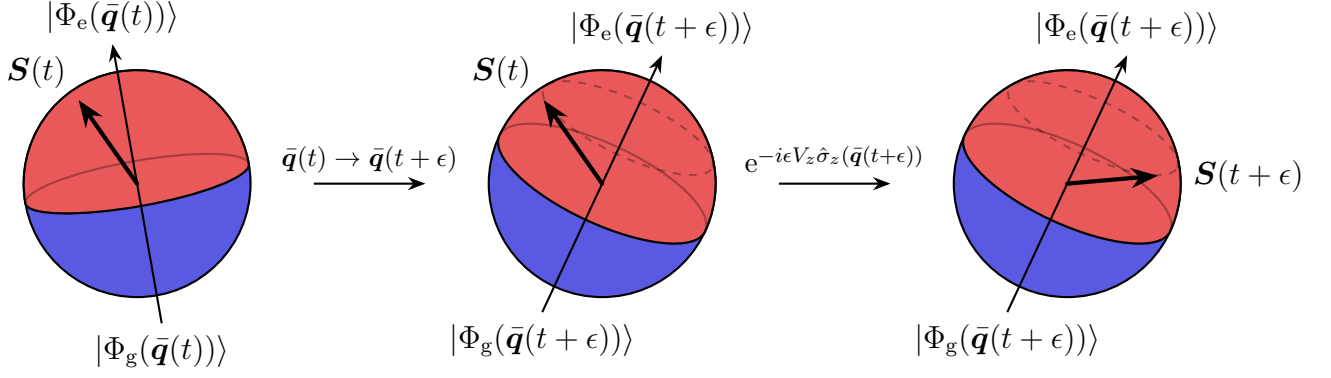
\begin{figure}
    \begin{tikzpicture}[scale=1.5, >=Stealth]

\definecolor{redhemi}{RGB}{230,60,60}
\definecolor{bluehemi}{RGB}{60,60,220}

\newcommand\anglebloch{10}
\newcommand\perspbloch{0.20}

\filldraw[fill=bluehemi!85, draw=black, thick]
    (-1,0) arc (180:-180:1)   %
    -- cycle;

\filldraw[fill=redhemi!85, draw=black, thick, rotate around={\anglebloch:(0,0)}]
    (-1,0) arc (180:360:1 and \perspbloch)  %
    arc (0:180:1)                    %
    -- cycle;

\draw[thick, opacity=0.3, rotate around={\anglebloch:(0,0)}] (0,0) ellipse (1 and \perspbloch);

\draw[->, thick, rotate around={\anglebloch:(0,0)}] (0,0) -- (0,1.25) node[above] {$\ket{\Phi_\text{e}(\Bq(t))}$};
\draw[-, thick, rotate around={\anglebloch:(0,0)}] (0,0) -- (0,-1.25) node[below] {$\ket{\Phi_\text{g}(\Bq(t))}$};

\draw[->, ultra thick, rotate around={35:(0,0)}, cap=round] (0,0) -- (0,0.9) node[above left] {$\bm{S}(t)$};

\newcommand\offsetbloch{4.1}
\renewcommand\anglebloch{-25}
\renewcommand\perspbloch{0.35}

\draw[->, thick] (0.38*\offsetbloch,0) -- (0.62*\offsetbloch,0) node[midway, above] {\footnotesize${\Bq(t)} \to {\Bq(t+\epsilon)}$};

\filldraw[fill=bluehemi!85, draw=black, thick, xshift=\offsetbloch cm]
    (-1,0) arc (180:-180:1)   %
    -- cycle;

\filldraw[fill=redhemi!85, draw=black, thick, xshift=\offsetbloch cm, rotate around={\anglebloch:(0,0)}]
    (-1,0) arc (180:360:1 and \perspbloch)  %
    arc (0:180:1)                    %
    -- cycle;

\draw[thick, opacity=0.3, xshift=\offsetbloch cm, rotate around={\anglebloch:(0,0)}] (0,0) ellipse (1 and \perspbloch);

\draw[->, thick, xshift=\offsetbloch cm, rotate around={\anglebloch:(0,0)}] (0,0) -- (0,1.25) node[above] {$\ket{\Phi_\text{e}(\Bq(t+\epsilon))}$};
\draw[-, thick, xshift=\offsetbloch cm, rotate around={\anglebloch:(0,0)}] (0,0) -- (0,-1.25) node[below] {$\ket{\Phi_\text{g}(\Bq(t+\epsilon))}$};

\draw[->, ultra thick, xshift=\offsetbloch cm, rotate around={35:(0,0)}, cap=round] 
(0,0) -- (0,0.9) node[above left] {$\bm{S}(t)$};

\draw[-, thick, dashed, xshift=\offsetbloch cm, rotate around={\anglebloch:(0,0)}, yshift=0.54 cm, opacity=0.3] 
    (0,0) ellipse (0.82 and 0.25)  %
    -- cycle;

\draw[->, thick] (1.38*\offsetbloch,0) -- (1.62*\offsetbloch,0) node[midway, above] {\small$\mathrm{e}^{-i \epsilon V_z \hat \sigma_z(\Bq(t+\epsilon))}$};

\filldraw[fill=bluehemi!85, draw=black, thick, xshift=2*\offsetbloch cm]
    (-1,0) arc (180:-180:1)   %
    -- cycle;

\filldraw[fill=redhemi!85, draw=black, thick, xshift=2*\offsetbloch cm, rotate around={\anglebloch:(0,0)}]
    (-1,0) arc (180:360:1 and \perspbloch)  %
    arc (0:180:1)                    %
    -- cycle;

\draw[thick, opacity=0.3, xshift=2*\offsetbloch cm, rotate around={\anglebloch:(0,0)}] (0,0) ellipse (1 and \perspbloch);

\draw[->, thick, xshift=2*\offsetbloch cm, rotate around={\anglebloch:(0,0)}] (0,0) -- (0,1.25) node[above] {$\ket{\Phi_\text{e}(\Bq(t+\epsilon))}$};
\draw[-, thick, xshift=2*\offsetbloch cm, rotate around={\anglebloch:(0,0)}] (0,0) -- (0,-1.25) node[below] {$\ket{\Phi_\text{g}(\Bq(t+\epsilon))}$};

\draw[-, thick, dashed, xshift=2*\offsetbloch cm, rotate around={\anglebloch:(0,0)}, yshift=0.54 cm, opacity=0.3] 
    (0,0) ellipse (0.82 and 0.25)  %
    -- cycle;

\draw[->, ultra thick, xshift=2*\offsetbloch cm, cap=round] 
(0,0) -- (0.8,0.07) node[right = 0.3cm] {$\bm{S}(t + \epsilon)$};

\end{tikzpicture}
    \centering
    \caption{Time evolution of the spin vector $\S$ induced by \eq{eq:winv_prop}. It can be decomposed into two contributions: a rotation around the $y$-axis (out-of-plane) from the changing character of the adiabatic states; and a precession around the new $z$-axis on the dashed circle of latitude resulting from the energy difference between the adiabatic states, $V_z$.
    }
    \label{fig:sphere}
\end{figure}

In particular, we will be interested in the following unitary transform corresponding to the propagation of the inverse kernel along a nuclear path $\Bq(t)$ according to the electronic Hamiltonian:
\begin{align} \label{eq:winv_prop}
    \hat w_{\Bq(t+\epsilon),\text{lin}}^{-1}\bk[big]{\S(t+\epsilon)} &=
    \eu{-i\epsilon\hat V(\Bq(t+\epsilon))} \, \hat w_{\Bq(t),\text{lin}}^{-1}\bk[big]{\S(t)} \, \eu{i\epsilon\hat V(\Bq(t+\epsilon))}.
\end{align}
Note that in addition to the rotation induced by the potential operator, there is a basis transform from the adiabatic states of the old position, $\Bq(t)$, to those of the new position, $\Bq(t+\epsilon)$, as illustrated in Fig.~\ref{fig:sphere}.

To obtain the equation of motion of $\S(t)$, we differentiate both sides of Eq.~\eqref{eq:winv_prop} with respect to $\epsilon$ using the chain rule and take the limit $\epsilon\rightarrow0$.
This gives
\begin{align}
    \sum_{\mu\in\{x,y,z\}} \dot{S}_\mu\pauli_\mu(\Bq) + \dot{\Bq}\cdot\nabla \winv_{\Bq,\text{lin}}(\S) = -i\left[\hat{V}(\Bq), \winv_{\Bq,\text{lin}}(\S)\right],
\end{align}
where $\nabla \winv_{\Bq,\text{lin}}(\S) = i \nac(\Bq) \sbk{\hat \sigma_y(\Bq), \winv_{\Bq,\text{lin}}(\S)}$,
and $\sbk{\hat{V}(\Bq), \winv_{\Bq,\text{lin}}(\S)} = V_z(\Bq)\sbk{\pauli_z(\Bq), \winv_{\Bq,\text{lin}}(\S)}$ because the contribution from the scalar potential $V_0(\Bq)$ cancels in the commutator.
Finally, %
projecting out individual values of the time derivatives $\dot{S}_\mu$ by multiplying by Pauli operators and taking the trace,
we obtain the equation of motion for the spin vector, $\S$:
\begin{align} \label{spinEOM}
    \dot \S
    &= 
    \begin{pmatrix} 0 \\ 2 \dot{\Bq} \cdot \nac(\Bq) \\ 2 V_z(\Bq)
    \end{pmatrix} \times \S.
\end{align}
The equation of motion for $\S'$ is equivalent with no change of sign.
The spin dynamics are thus identical to those of the original MASH method\cite{MASH} and other spin-mapping approaches,\cite{spinmap,spinPLDM1,spinPLDM2}
which precess around an effective magnetic field determined by the energy gap and the nonadiabatic coupling vector.

Before finishing this section,
we note that the inverse kernel introduced above is not unique.
In particular, we have found the following variants to provide useful alternatives, which we explore in more depth in Sec.~\ref{sec:results}:
\begin{subequations} \label{eq:variants}
\begin{align}
    \hat w_{\q,\mathrm{init}}^{-1}(\S) & %
    = 2\sbk{\sum_{\mu\in\{x,y,z\}} \hat{\sigma}_\mu(\q) \cdot S_\mu} \hat h_{\q}(\S),
    \label{eq:initial_kernel}
    \\
    \hat w_{\q,\mathrm{fin}}^{-1}(\S) &  %
    = 2\hat h_{\q}(\S) \sbk{\sum_{\mu\in\{x,y,z\}} \hat{\sigma}_\mu(\q) \cdot S_\mu},\label{eq:final_kernel}
\end{align}
\end{subequations}
where $\hat h_{\q}(\S) = \ketbra{\Phi_\mathrm{g}(\q)}{\Phi_\mathrm{g}(\q)} h(-S_z) + \ketbra{\Phi_\mathrm{e}(\q)}{\Phi_\mathrm{e}(\q)} h(S_z)$.
Note that all the defining properties of a valid inverse mapping kernel are also reproduced by these alternative versions.
However, as the derivation of the MASH-PLDM method is slightly more complicated in this case, we will continue to focus on the use of \eqn{eq:MASH_PLDM_inverse} in the main text and describe the necessary extensions relevant for the variants in Appendix~\ref{app:variant}.

\subsection{Derivation of MASH-PLDM} \label{sec:MASHPLDM}

In this subsection, we use the MASH kernels defined in Sec.~\ref{sec:mapping} to map the path-integral representation of the QCLE from Sec.~\ref{sec:PLDM} and thus derive the MASH-PLDM method.
The goal is to apply the mapping procedure to the operators in \eqn{eq:hatT} that appear in combination with $\Dq_k$, namely $\nabla \hat V(\Bq_k)$, in order to relieve them from their operator nature and enable integration over $\Dq_k$. The mapping procedure itself is exact and thus, if we were to map all these operators individually and integrate over $\Dq_k$, it would result in an expression equivalent to the QCLE but expressed in terms of $2N$ integrals over separate Bloch spheres, which would be no more tractable than the original expression.
However, as we show in Appendix~\ref{app:perturbation} with a perturbative approximation, it is sufficient to introduce only one set of mapping variables for each of the forward and backward paths, and by making use of the property introduced in Eq.~\eqref{eq:winv_prop}, the inverse kernel can be moved to the end of the expression. %
From that we obtain 
\begin{align}
    \hat{T} \simeq
    \int\rmd \S\, \eu{-\frac{i\epsilon}{2} \nabla V\bk{\Bq_{N}, \S(N\epsilon)}\cdot\Dq_{N}} \, \eu{-i\epsilon\hat{V}(\Bq_{N})}
    \cdots \eu{-\frac{i\epsilon}{2} \nabla V\bk{\Bq_{1}, \S(\epsilon)}\cdot\Dq_1} \, \eu{-i\epsilon\hat{V}(\Bq_1)} \, \hat{w}_{\Bq_0}^{-1}(\S),
\end{align}
where $\nabla V(\Bq_{k}, \S(k\epsilon)) = \tr\sbk{\nabla\hat V(\Bq_k)\, \w_{\Bq_k}(\S(k\epsilon))}$
and the time-evolved spins $\S(k\epsilon)$ are defined through their equation of motion \eqn{spinEOM}.
The analogous expression for $\hat{T}'$ is defined in terms of the integral over the mapping variable $\S'$.
As we shall show, this approximation leads to a practical trajectory-based method for simulating nonadiabatic dynamics.
Although it is only formally equivalent to the QCLE in the perturbative limit,
the system--bath coupling is treated to all orders at least within a quasiclassical approximation, making the method more powerful than typical perturbative master equations.\cite{OpenQuantum}

By inserting this approximation for $\hat{T}$ and $\hat{T}'$ in Eq.~\eqref{eq:QCLE},
we can now collect the phase terms
such that instead of just $\eu{i\Delta\mathcal{S}_\mathrm{n}}$, we now have a phase-factor of $\eu{i(\Delta \mathcal{S}_\mathrm{n} + \mathcal{S}_\mathrm{e} - \mathcal{S}_\mathrm{e}')}$,
where $\Delta\mathcal{S}_\mathrm{n}$ was given in Eq.~\eqref{eq:DeltaSn} and the new contributions to the phase are
\begin{equation}
    \mathcal{S}_\mathrm{e}^{(\prime)} = \pm \frac\epsilon2 \sum_{k=1}^{N} \bm{F}\bk[big]{\Bq_k,\S^{(\prime)}(k\epsilon)} \cdot \Dq_k,
\end{equation}
where the new phase terms include a mapped force function defined by
\begin{equation} \label{eq:MASHforce}
    \bm{F}(\Bq,\S) = - \nabla V_0(\Bq) - \nabla V_z(\Bq) \sgn(S_{z}) + 4 \nac(\Bq) V_z(\Bq) \delta(S_{z})S_{x} ,
\end{equation}
which is recognized as the MASH force from \Refx{MASH}.

Finally, integrating the phase factor over the difference coordinates, $\Dq_k$ and $\Dp_k$, results in Dirac delta functions
that %
determine the equations of motion for the nuclei at time $t_k = k \epsilon$ in terms of the forces.
In the continuous representation, these are
\begin{subequations}\label{nuclearEOM}
    \begin{align} 
    \frac{\d \Bq}{\d t} &= \frac{\Bp}{m},\\
    \frac{\d \Bp}{\d t} &= \half\big[\bm{F}(\Bq,\S) + \bm{F}(\Bq,\S')\big] . \label{force}
    \end{align}
\end{subequations}

Putting everything together, we obtain the final working equation for MASH-PLDM:
\begin{align} \label{eq:MASH-PLDM}
    C_{AB}^\text{MASH-PLDM}(t) &= \int \frac{\d\Bq\,\d\Bp}{(2\pi)^f}\,\d\S\,\d\S' \, \rho_\text{n}(\Bq, \Bp) \tr \left[\hat A \, \winv_{\Bq,\mathrm{lin}}(\S') \, \hat{U}^\dagger \, \hat B(\Bq(t),\Bp(t)) \, \hat{U} \, \winv_{\Bq,\mathrm{lin}}(\S)\right],
\end{align}
where we renamed $\Bq = \Bq_0$ and $\Bp = \Bp_0$ and defined the unitary evolution operator
$\hat{U} = \eu{-i\epsilon\hat V(\Bq_N)}\cdots\eu{-i\epsilon\hat V(\Bq_1)}$.
To evaluate the expression in practice, one samples nuclear positions and momenta from a Wigner distribution and two sets of mapping variables uniformly from the Bloch sphere.
Starting from these initial conditions, the dynamical variables are propagated using Eqs.~\eqref{nuclearEOM} and \eqref{spinEOM} and the observables are measured via the trace over the electronic subspace.

Due to the similarity of the formalism, many considerations valid for the original MASH method can be directly transferred to MASH-PLDM\@.
For instance,
the nuclear force defined in \eq{force} is simply the average of two MASH-like forces, %
and because MASH trajectories are energy conserving by design,
MASH-PLDM trajectories also conserve an energy, given by %
\begin{align} \label{eq:E}
    E(\Bq,\Bp,\S,\S') = \frac{||\Bp||^2}{2m} + V_0(\Bq) + \half\big[\sgn(S_z) + \sgn(S_z')\big] V_z(\Bq).
\end{align}

Additionally, MASH-PLDM has a momentum-rescaling procedure similar to that of MASH,\cite{MASH}
in which an impulse is applied to the nuclear momentum when a spin vector crosses the equator. %
In particular, the new momentum is determined by
\begin{equation}\label{momentum_rescaling}
    \frac{\tilde p_\mathrm{new}^2}{2m} = \frac{\tilde p_\mathrm{old}^2}{2m} - \sgn(\dot S_z) \, V_z(\Bq),
\end{equation} 
where $\tilde p_\mathrm{old/new} = \Bp_\mathrm{old/new} \cdot \nac(\Bq) / \|{\nac(\Bq)}\|$ are the projections of the momentum before/after the hop onto the direction of the nonadiabatic coupling vector (all mass weighted). The sign of $\dot S_z$ indicates whether the spin is hopping up or down; thus the subtracted term corresponds to the energy difference of hopping to or from the average potential energy surface (and hence the factor of 2 difference from the similar equation in \Refx{MASH}). In the case where the total right-hand side of \eq{momentum_rescaling} is negative, the available kinetic energy is insufficient to excite the system to the upper electronic state and a frustrated hop occurs.  In this case, the spin reflects off the equator ($\dot S_{z, \mathrm{new}} = -\dot S_{z, \mathrm{old}}$) and $\tilde p$ is elastically reflected ($\tilde p_\mathrm{new} = - \tilde p_\mathrm{old}$). 
This is the procedure that is naturally obtained from the derivation and is equivalent to that of the original FSSH\cite{HammesSchiffer1994FSSH} and MASH\cite{MASH} methods.  However, a number of alternatives have been proposed,\cite{Mueller1997FSSH,Jasper2003FSSH,piMASH} which could be explored in future work.

Furthermore, we used a modified velocity-Verlet integration scheme developed for MASH\cite{MASHEOM} to propagate the coupled equations of motion for the nuclei and spins.
It exhibits a convergence rate of $\mathcal{O}(\epsilon^2)$ and is time-reversible thanks to a linear bisection algorithm for the sub-timestep evaluation of the hopping times.
Note that for simplicity, the derivation presented above uses the most basic form of Trotter splitting for the evolution operator $\hat U$. %
However, it is more efficient to use the symmetric time-ordered product,
$\hat{U}=\eu{-i\epsilon\hat{V}(\Bq_N)/2} \, \eu{-i\epsilon\hat{V}(\Bq_{N-1})} \cdots \eu{-i\epsilon\hat{V}(\Bq_1)} \, \eu{-i\epsilon\hat{V}(\Bq_{0})/2}$, which accelerates convergence with respect to the timestep.
In our implementation, the timesteps added by the bisection algorithm were also explicitly included in the evolution operator $\hat U$, although this is not expected to have a significant effect on the convergence behavior.

In order to evolve the spins and construct the evolution operator, one needs to take into account the fact that the basis functions change at each nuclear configuration along the trajectory (Fig.~\ref{fig:sphere}).
In this work, we use the nonadiabatic coupling vector to propagate the spin vector as in \eqn{spinEOM}.
However, following many surface-hopping studies,\cite{HammesSchiffer1994FSSH,Plasser2012FSSH,Meek2014overlaps,Mai2018SHARC}
one could consider using wavefunction overlaps instead.\cite{MASHEOM}
Likewise, for the construction of the evolution operator, we employed the nonadiabatic coupling vectors to perform the basis rotation, %
in particular by using
$\braket{\Phi_\ell(\Bq_{k+1})|\eu{-i\epsilon\hat{V}(\Bq_k)}|\Phi_{\ell'}(\Bq_k)}
\simeq
\braket{\Phi_\ell(\Bq_{k})|\eu{-i\epsilon\hat{V}_\text{eff}(\Bq_k)}|\Phi_{\ell'}(\Bq_k)}$,
where the effective potential is
$\hat{V}_\text{eff}(\Bq_k) = V_z(\Bq_k)\pauli_z(\Bq_k) + \frac{\nac(\Bq_k)\cdot\Bp_k}{m}\pauli_y(\Bq_k)$.
Alternatively, one could use wavefunction overlaps to transform the basis or construct the operator in the diabatic representation, in which the $\hat{V}$ operators are not diagonal.

\subsection{Analysis of MASH-PLDM}\label{sec:analysis}

\begin{figure}[h]
    \includegraphics[width=0.5\textwidth]{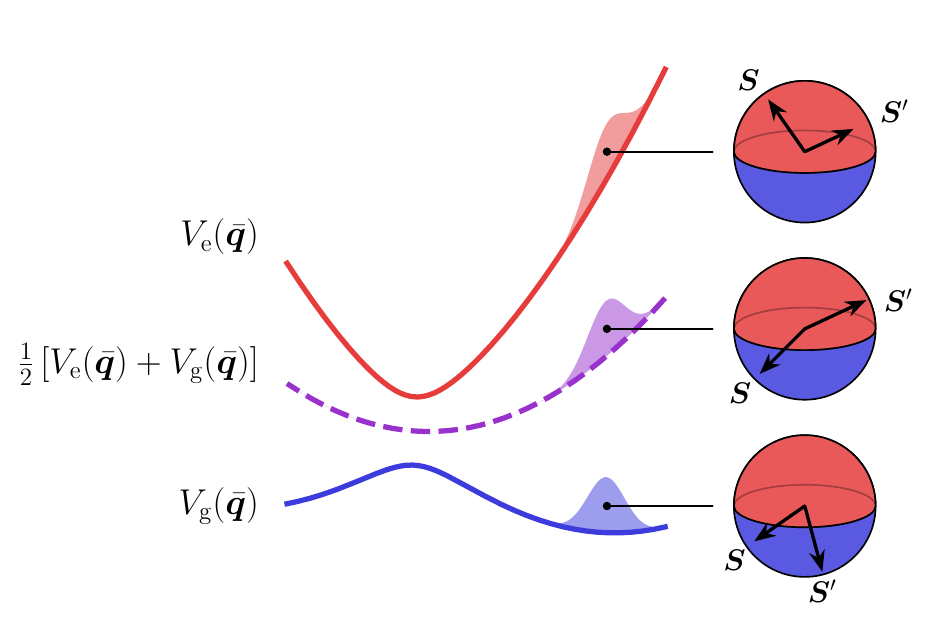}
    \centering
    \caption{%
    In MASH-PLDM, the nuclear force is determined based on the hemispheres in which the spins $\S$ and $\S'$ are located.
    The force is that of the excited state if both spins are in the upper hemisphere, of the lower state if both are in the lower hemisphere, or averaged if they are in different hemispheres.
    }
    \label{fig:forces}
\end{figure}

One important difference between the original MASH method and the MASH-PLDM approach that we have just derived
is that each MASH-PLDM trajectory has two spin vectors rather than just one.
The spin vector in the original MASH method determines the active state depending on its current location on the Bloch sphere. %
Similarly, according to the MASH-PLDM force [\eqn{force}], the nuclei will move on the excited surface when both spin vectors are in the upper hemisphere, or on the ground surface when both are in the lower hemisphere.
However, in the case that one spin vector points up and the other down, the nuclei will feel a force which is averaged (in equal proportions) between the two states, as illustrated in Fig.~\ref{fig:forces}.

Allowing the nuclei to move on an average surface may sound like we have reintroduced the problems of mean-field theories.
However, it is known %
that propagating classical trajectories in this way is an excellent approximation when simulating coherences.
In fact, it forms the backbone of the WACL (Wigner averaged classical limit) method used for linear and nonlinear spectroscopy\cite{MukamelBook,Rabani1998vibronic,Egorov1998vibronic,Shi2005nonadiabatic,Shi2008nonlinear,McRobbie2009nonlinear,Karsten2018vibronic,nonlinear}
and has been identified as a desired property to improve the description of coherences in the original MASH method.\cite{Mannouch2024coherence}
Similar ideas have also been incorporated into other quasiclassical methods.\cite{vanderVegte2013nonlinear,Kapral2015QCL,Wang2015Liouville,Polley2020vibronic}

In fact, it is possible to prove that MASH-PLDM reduces to the WACL approach in the limiting case that the ground and excited states are decoupled.
To show this, we note that $S_z$ and $S_z'$ are constant in the absence of nonadiabatic coupling, such that the trajectories are dependent only on the hemispheres but not on the other details of the spin vectors.
Next, we note that
$\int \d\S \,\winv_{\Bq}(\S) h(S_z) = \ketbra{\Phi_\mathrm{e}(\Bq)}{\Phi_\mathrm{e}(\Bq)}$
and
$\int \d\S \, \winv_{\Bq}(\S) h(-S_z) = \ketbra{\Phi_\mathrm{g}(\Bq)}{\Phi_\mathrm{g}(\Bq)}$ for any valid inverse kernel.
This implies that if $\hat{A}$ is a population operator, both $\S$ and $\S'$ can be sampled from the appropriate hemisphere, as any other combination will average to zero, and the force will be determined from the appropriate active state.
However, when simulating the dynamics of a coherence such as $\hat{A}=\ketbra{\Phi_\mathrm{g}}{\Phi_\mathrm{e}}$, as is common for optical spectroscopy, the only nonzero contribution is found when $\S$ is sampled from the lower hemisphere and $\S'$ from the upper, such that the nuclei evolve on the average surface.
In addition, as $\hat{U}$ is diagonal in this simple case, it recovers the phase factor of the WACL approach, which depends on the cumulative integral of the energy gap.

In more general cases, however,
where there is nonadiabatic coupling between the states,
the analysis is more complicated.
We find that in general, there can be non-zero contributions from trajectories starting or ending with the spin vectors in the ``wrong'' hemispheres,
and as they move according to forces that are inconsistent with the electronic operators being measured,
the nuclear dynamics cannot be physical.
For this reason, we have developed variants of the kernels that include Heaviside step functions to select the physically relevant trajectories.
This is discussed further in Sec.~\ref{sec:results} and Appendix~\ref{app:variant}.

The derivation we have presented, which is justified through detailed proofs in Appendices \ref{app:perturbation}--\ref{app:QCLE},
shows that MASH-PLDM recovers (a version of) the QCLE dynamics to first-order in time. %
This is an important theoretical advantage over methods such as Ehrenfest and FSSH, which do not have this rigorous connection to the QCLE\@.\footnote{Under the Ehrenfest approximation, the correlation function is
\[ C_{AB}^\text{Ehrenfest}(t) = \int \frac{\d\q\,\d\p}{(2\pi)^f} \,\rho_\mathrm{n}(\q,\p) \, \tr\left[\hat{\rho}_\mathrm{el}(t) \hat{B}(\q(t),\p(t))\right] , \]
where $\hat{\rho}_\mathrm{el}(0) = \hat{A}$ is the electronic density matrix of a pure state,
and the equations of motion are
$\dot{\q}=\frac{\p}{m}$, $\dot{\p} = \tr[\hat{\rho}_\mathrm{el}(t) \hat{\bm{F}}]$ and
$\der{}{t}\hat{\rho}_\mathrm{el} = -i\left[\hat{V} + 2\frac{\nac\cdot\p}{m}\pauli_y, \hat{\rho}_\mathrm{el}\right]$.
Consider for example the time-derivative of the correlation function with $\hat{A}=\ketbra{\Phi_\mathrm{e}}{\Phi_\mathrm{e}}$ and $\hat{B}=B_x(\hat{\p})\otimes\pauli_x(\hat{\q})$:
\[
	\dot{C}_{AB}^\text{Ehrenfest} = \int \frac{\d\q\,\d\p}{(2\pi)^f}\,\rho_\mathrm{n}(\q,\p) \, \left\{ \tr\left[\der{\hat{\rho}_\mathrm{el}}{t} \hat{B}(\q(t),\p(t))\right]
	+ \tr\left[\hat{\rho}_\mathrm{el} \pder{\hat{B}}{\p} \cdot \dot{\p}\right]
	\right\} .
\]
The second term gives 0,
although 
according to QCLE it should be equal to
${\tr[\hat{\rho}_\mathrm{el} \half\pder{B_x}{\p}\cdot[\pauli_x,\hat{\bm{F}}]_+]} = 
\pder{B_x}{\p}\cdot 2\nac V_z$. %
Similarly, this term is missing under FSSH dynamics, which are in this case equivalent to Ehrenfest, as no hops out of an initial population can occur to first-order in time.}
The same claim has been made for FBTS,\cite{Hsieh2013FBTS}
spin-PLDM,\cite{spinPLDM2} and MASH,\cite{MASH,piMASH,MASHreview}
and an iterative quantum-jump procedure was proposed in each case which would systematically recover the full QCLE result, at the price of significantly increasing the computational cost.%
\footnote{A related iterative scheme was also proposed for the original PLDM method.\cite{Huo2012PLDM}
However, although it was observed to improve results, it does not rigorously recover the full QCLE\cite{Hsieh2012FBTS}}
In principle, one could apply the quantum-jump procedure to MASH-PLDM in a similar way.
However, our hope is that the improved dynamics of MASH-PLDM will render such corrections unnecessary and allow simulations to give accurate results within a reasonable computation time.
Therefore, all results presented in Sec.~\ref{sec:results} were computed without quantum jumps.

Because of the derivation based on the new mapping kernels,
the MASH-PLDM trajectories automatically conserve energy through the momentum rescaling that occurs at hops.
This is similar to the original MASH method, %
but stands in stark contrast to other nonadiabatic dynamics methods which must either introduce such rescalings into the method in an \emph{ad hoc} fashion,\cite{Tully1990hopping,wu2025NaF}
or live with the fact that their trajectories do not conserve energy.\cite{Martens2019QTSH,Arribas2023,Ibele2026}
In MASH, the fact that its trajectories conserve energy forms part of the argument that it will thermalize to the correct quantum--classical equilibrium distribution in an ergodic system.\cite{thermalization,MASHreview}
Unfortunately, this argument does not naturally generalize to the MASH-PLDM method, %
as the $\S$ and $\S'$ vectors do not move independently of each other
(and are not easily integrated out due to the nonlinearity of the dynamics)
such that the ergodic hypothesis cannot be invoked.
Nonetheless, the results shown in Sec.~\ref{sec:spinboson} seem to indicate that MASH-PLDM has similar thermalization behavior to MASH.

The quantum-mechanical correlation function $C_{AB}(t)$ is known to be purely real in the case that $\hat{A}$ and $\hat{B}$ are Hermitian operators.
It is easy to show that MASH-PLDM obeys this property once the integrals are converged.
In particular, taking the Hermitian conjugate of the operators inside the trace of \eqn{eq:MASH-PLDM} effectively reverses their order and thus just exchanges the dummy variables $\S$ and $\S'$.
For this reason, we only present the real part of our computed correlation functions, as one can treat the imaginary part purely as statistical error, which will disappear in the limit of infinite sampling.

Finally, we note that although the mapping kernel recovers the MASH force, it does not map observables in the same way as those in the original MASH method.
One might, therefore, ask whether an alternative fully-linearized MASH method can be defined based on these kernels.
We study this question in Appendix~\ref{app:MASHLSC} %
and find that it is possible to construct correlation functions based on these kernels which formally also reduce to the QCLE (in certain limits) as for the original MASH method.
However, this new approach is not as practical as the original formalism, as delta functions appear in some observables.
Nonetheless, these issues 
are of no consequence for the use of these kernels in MASH-PLDM, where we are not required to map the observables at all.

\section{Results} \label{sec:results}

In this section, we present results for a set of model systems which have been extensively used to benchmark a wide range of nonadiabatic dynamics methods.
To avoid cluttering our plots, we focus in particular on the differences between MASH, spin-PLDM and MASH-PLDM\@.
Results from other methods can be compared using the references to their original literature.

\subsection{Scattering Models}\label{sec:scattering_models}

First, we test the one-dimensional ($f=1$) scattering models introduced by Tully.\cite{Tully1990hopping}
The Hamiltonians in the diabatic representation are
\begin{align} \label{Hdia}
    \hat{H} &= \frac{p^2}{2m} +  \begin{pmatrix}
        V_{11}(q) & V_{12}(q) \\
        V_{12}(q) & V_{22}(q)
    \end{pmatrix}.
\end{align}
In each case, we work in atomic units with a nuclear mass of $m = \num{2000}$.
Note that MASH-PLDM dynamics are carried out in the adiabatic representation, which is generated by diagonalizing the potential energy matrix.

Depending on whether we simulate the time evolution of a wavepacket 
or an energy-resolved scattering experiment,
the nuclei are initialized by %
\begin{subequations}
\begin{align}
    \rho_\mathrm{n}(q, p) &= 2 \exp\sbk{-\frac{1}{\gamma_0} (p-p_\text{init})^2 - \gamma_0 (q - q_\text{init})^2} ,
    \label{eq:nucl_wavepacket_tully}
\intertext{where $\gamma_0=\tfrac12$, or} %
    \rho_\mathrm{n}(q,p) &= 2\pi \delta(q-q_\text{init}) \delta(p - p_\text{init}),
    \label{eq:nucl_deltas_tully}
\end{align}
\end{subequations}
respectively. In each case, the correlation function is measured starting from the electronic ground state $\hat A = \ketbra{\Phi_\mathrm{g}}{\Phi_\mathrm{g}}$
and we introduce the alternative notation $\braket{\hat{B}(t)} = C_{AB}(t)$ and $\hat{P}_\mathrm{e}=\ketbra{\Phi_\mathrm{e}}{\Phi_\mathrm{e}}$.
Note that with $q_\text{init} = -15$ in each case, the nuclei are initialized far enough to the left of the nonadiabatic coupling region, so that the adiabatic initial condition is essentially equivalent to a diabatic initial condition. Energy-resolved scattering calculations were run until the trajectory left the scattering region, i.e., $\bar{q}(t) \notin [-15, 15]$. 
The spin-PLDM results presented in this section use focused initial sampling, as described in \Refx{spinPLDM2}.

Most results presented in this section are obtained from 10 million randomly sampled trajectories with a timestep of $\epsilon = 1$.
However, for energy-resolved scattering experiments using MASH-PLDM, a Sobol sequence\cite{sobol} was used to generate a quasi-random distribution of spin vectors to accelerate convergence, reducing the number of samples to approximately one million ($2^{20}$). %
In many cases, far fewer trajectories and larger timesteps will be sufficient; these parameters were selected to completely eliminate statistical and timestep errors to graphical accuracy in order to compare the methods at their theoretical best. We refer the reader to the supplementary material where the convergence rate of MASH-PLDM is compared to that of MASH, in which it is seen to require only about a factor of 3--4 more trajectories for the same desired statistical error (when using the kernel with initial projections).
This means that reasonable results (where the enhanced accuracy of MASH-PLDM over MASH becomes apparent) can already be obtained with a few thousand trajectories.

Exact quantum benchmarks are generated by solving the Schr\"odinger equation in the diabatic representation, either using the split-operator method\cite{Leforestier1991quantum} to propagate an initial wavepacket $\chi(q)=(\gamma_0/\pi)^{1/4}\,\eu{-\gamma_0 (q-q_\text{init})^2/2+ip_\text{init}(q-q_\text{init})}$ or the log-derivative method to solve for the stationary scattering states.\cite{Mano1986logderivative,Alexander1989logderivative}

\subsubsection{Tully model I} \label{sec:Tully1}
We employ a variant form of Tully's model I of a single avoided crossing with analytic potentials defined by \cite{Ananth2007SCIVR}
\begin{subequations}
\begin{align}
    V_{11}(q) = -V_{22}(q) &= a \tanh(bq),\\
    V_{12}(q) &= c\,\e^{-d q^2},
\end{align}
\end{subequations}
where $a = 0.01$, $b = 1.6$, $c = 0.005$, and $d = 1$.
A wavepacket is initialized on the left with an initial average kinetic energy of $E_0 = 0.03$ (i.e., $p_\text{init} \approx 10.95$).
The observable $\hat{B}=\delta(\bar{q}(t)-q_\text{fin})\otimes\Id$ %
is measured at $t=\SI{150}{fs}$.

\begin{figure}
    \centering
    \includegraphics[width=0.45\linewidth]{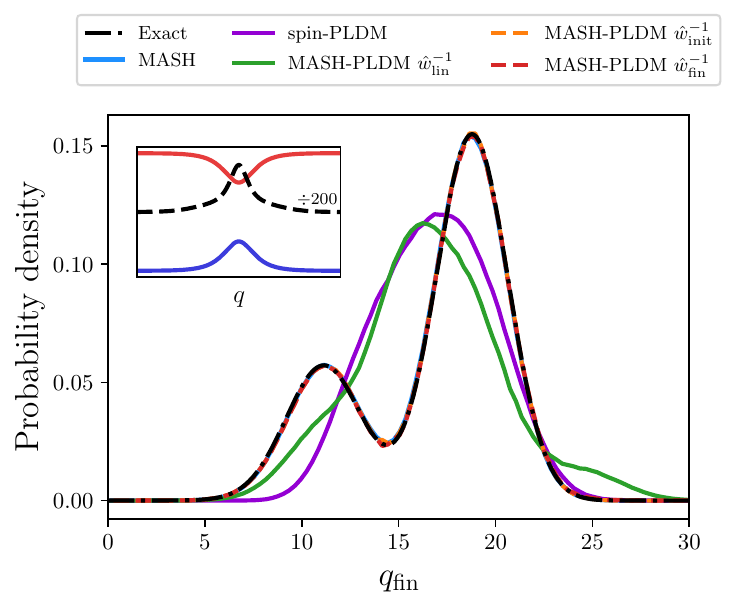}
    \caption{%
    Final position distribution after evolving for \SI{150}{fs} in Tully's model I\@.
    Both MASH and the MASH-PLDM variants agree well with the exact quantum benchmark except when using the linear inverse kernel, $\winv_\text{lin}$.
    The inset depicts the adiabatic potentials and nonadiabatic coupling vector scaled down by a factor of 200.
    }
    \label{fig:tully1_wavepacket}
\end{figure}

Results generated by MASH-PLDM are shown in Fig.~\ref{fig:tully1_wavepacket} and compared with the exact quantum benchmark, original MASH and spin-PLDM\@.
Spin-PLDM, like the original PLDM, and other methods based on mean-field mapping approaches,\cite{Kelly2012mapping,Cotton2013mapping,spinmap,linearized} fails to capture the wavepacket branching,
unless more expensive methods are employed (such as quantum jumps,\cite{Huo2012PLDM} semiclassical phases\cite{Ananth2007SCIVR,Miller2009mapping} or nonpositive-definite weighting functions\cite{SPMD}). %

Surface-hopping methods were developed for solving this problem in an inexpensive way.
It is therefore unsatisfying that MASH-PLDM with the linear inverse kernel [\eqn{eq:MASH_PLDM_inverse}]
is also unable to correctly describe the wavepacket splitting.
This is because trajectories moving on the ``wrong'' surfaces are contributing to the overall result.
In particular, even though the system is supposed to start in the electronic ground state, there are trajectories initialized on the upper and average potential energy surface in addition to the lower surface.
In addition, after passing through the avoided crossing, trajectories may end up on the average surface, even though they contribute towards the identity operator in $\hat{B}$.

These problems are resolved using the more powerful inverse kernels, such as \eqn{eq:initial_kernel}, which removes the contributions from trajectories initialized on the wrong surface.
With this approach, MASH-PLDM recovers the quantum result almost perfectly.
A similarly good result is obtained using \eqn{eq:final_kernel} to select the appropriate trajectories at the final time.

However, the original MASH method,\cite{MASH} and most other variants of surface hopping,\cite{Tully1990hopping} also obtain almost perfect results when simulating this simple system.
It is therefore necessary to turn to a more difficult example to illustrate the advantages of MASH-PLDM.

\subsubsection{Tully model II}
\label{sec:Tully2}

The diabatic states of Tully's model II of a dual avoided crossing are defined by
\begin{subequations}
\begin{align}
    V_{11}(q) &= 0 , \\
    V_{22}(q) &= - a\,\e^{-b q^2} + \varepsilon,\\
    V_{12}(q) &= c\,\e^{-d q^2},
\end{align}
\end{subequations}
where $a=0.1$, $b=0.28$,  $c = 0.015$, $d = 0.06$, and $\varepsilon = 0.05$.
We simulate a scattering event for a range of different initial momenta and calculate the transmission probability on the excited state with $\hat{B} = h(p(t))h(q(t)-15) \otimes \ketbra{\Phi_\mathrm{e}}{\Phi_\mathrm{e}}$.

When studying scattering problems, it is natural to think about the dynamics as a function of energy rather than as a function of initial momentum.
This has an important consequence for MASH-PLDM, for which trajectories may be initialized on the upper or average surface, even when the initial population is fully in the ground state.
We therefore employed the energy-based initial distribution
$\rho_\mathrm{n}(q,p) = 2\pi\delta(q-q_\text{init}) \delta(p - \tilde{p}_\text{init})$,
where the rescaled initial momentum $\tilde p_\text{init}$ is defined as the solution of $E(q_\text{init},\tilde{p}_\text{init},\S,\S') = p_\text{init}^2/2m + V_\mathrm{g}(q_\text{init})$ [where $E$ is given in \eqn{eq:E}].
In cases where there is no real-valued solution, %
no trajectory is run and 
there is no contribution to the correlation function.
A similar approach was employed in \Refx{SPMD}.
Note that this alternative initialization procedure has no effect on MASH-PLDM with the initial inverse kernel, as then all trajectories anyway start with both spins in the lower hemisphere.

\begin{figure}
    \centering
    \includegraphics[width=0.45\linewidth]{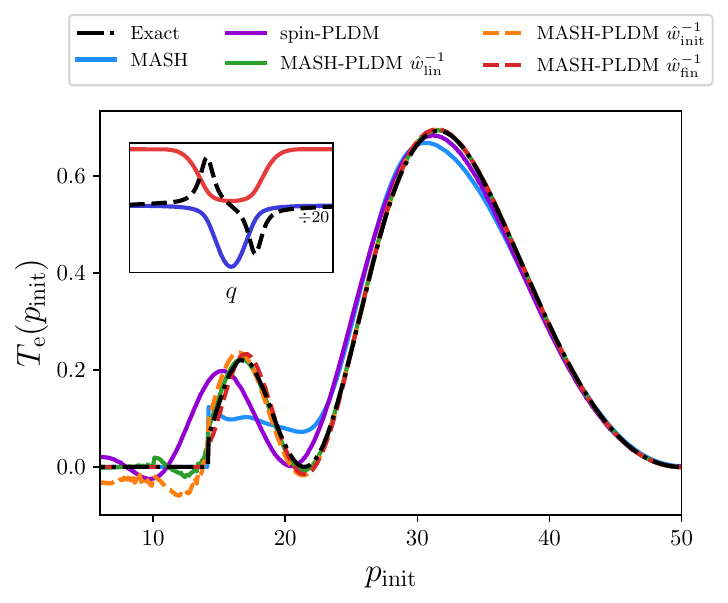}
    \caption{The scattering probability of transmission to the upper adiabat as a function of the initial momentum, $p_\text{init}$, for Tully's model II\@.
    The inset depicts the adiabatic potentials and nonadiabatic coupling vector.
    }
    \label{fig:tully2_scatter}
\end{figure}

Using this approach, the scattering results
are presented in Fig.~\ref{fig:tully2_scatter}.
The agreement with quantum mechanics is excellent,
and in particular fixes %
many of the errors of spin-PLDM and the original MASH method.
MASH-PLDM may, however, give negative probabilities at low energies.
This unphysical problem is eliminated by using the inverse kernel with a projection at the final time [\eqn{eq:MASH-PLDM_final}], which only allows trajectories that end with both spins in the upper hemisphere to contribute to the excited-state transmission. %
This, along with the %
energy-based initial distribution,
ensures that none of the trajectories in this low-energy region contribute, as they do not have enough energy to hop to the upper state.
As discussed in Sec.~\ref{sec:analysis}, this is not the case with the linear and initial kernels, where any trajectory may contribute to the excited-state population (either positively or negatively) regardless of their spin configuration.

With this best choice of kernel, the MASH-PLDM result is clearly superior to Ehrenfest, linearized spin-mapping, and FSSH, whose results can be seen in \Refx{MASH},
and similar if not slightly more accurate than the symmetrical windowing quasiclassical (SQC) approach.\cite{Cotton2013mapping}

\begin{figure}
    \centering
    \includegraphics[width=0.45\linewidth]{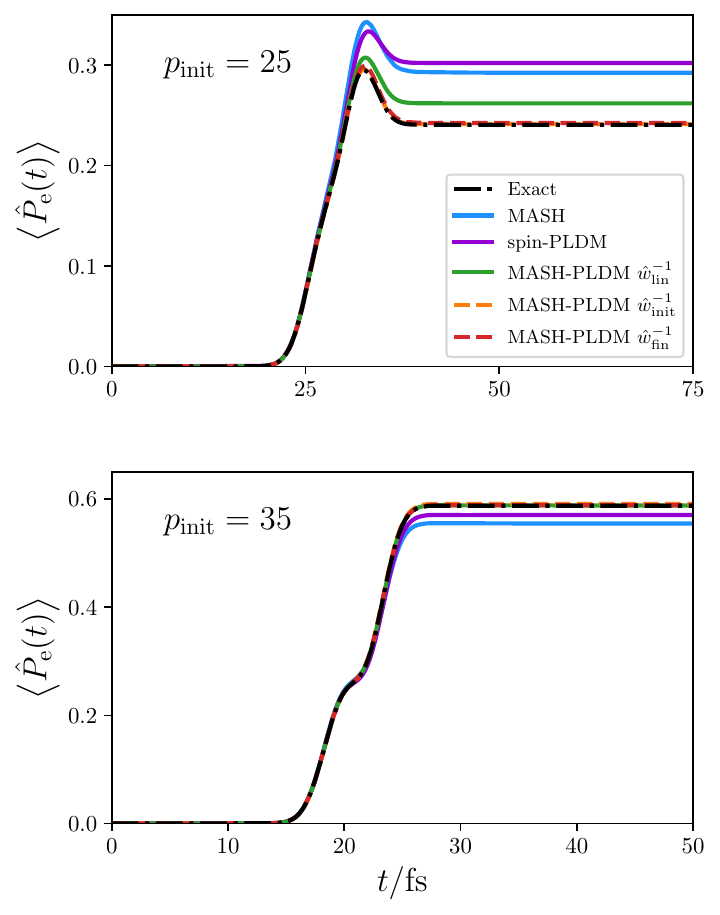}
    \caption{The time-dependent populations of the upper adiabatic state associated with the wavepacket dynamics of Tully's model II, with different average initial momenta, $p_\mathrm{init}$. %
    }
    \label{fig:tully2_pops}
\end{figure}

To further demonstrate the improvement over original MASH,
we additionally show time-dependent populations generated by an initial wavepacket in Fig.~\ref{fig:tully2_pops}.
Note that we do not employ the energy-based initial conditions here, as the result is explicitly time dependent.
The original MASH method %
required computationally expensive
quantum jumps %
in order to obtain the correct behavior for $p_\mathrm{init}=35$ and even after 4 jumps was still not perfect for $p_\mathrm{init}=25$.\cite{MASH}
MASH-PLDM (with either the initial or final inverse kernels), however, gives very accurate results far more efficiently without using quantum jumps.

\subsubsection{Tully model III}
\label{sec:Tully3}

Tully's model III of extended coupling with reflection is defined by
\begin{subequations}
\begin{align}
    V_{11}(q) &= - V_{22}(q) = a,\\
    V_{12}(q) &= b\sbk{1+ \sgn(q)(1-\e^{-c |q|})},
\end{align}
\end{subequations}
where $a={6\cdot10^{-4}}$, $b = 0.1$, and $c = 0.9$.

\begin{figure}
    \centering
    \includegraphics[width=0.7\linewidth]{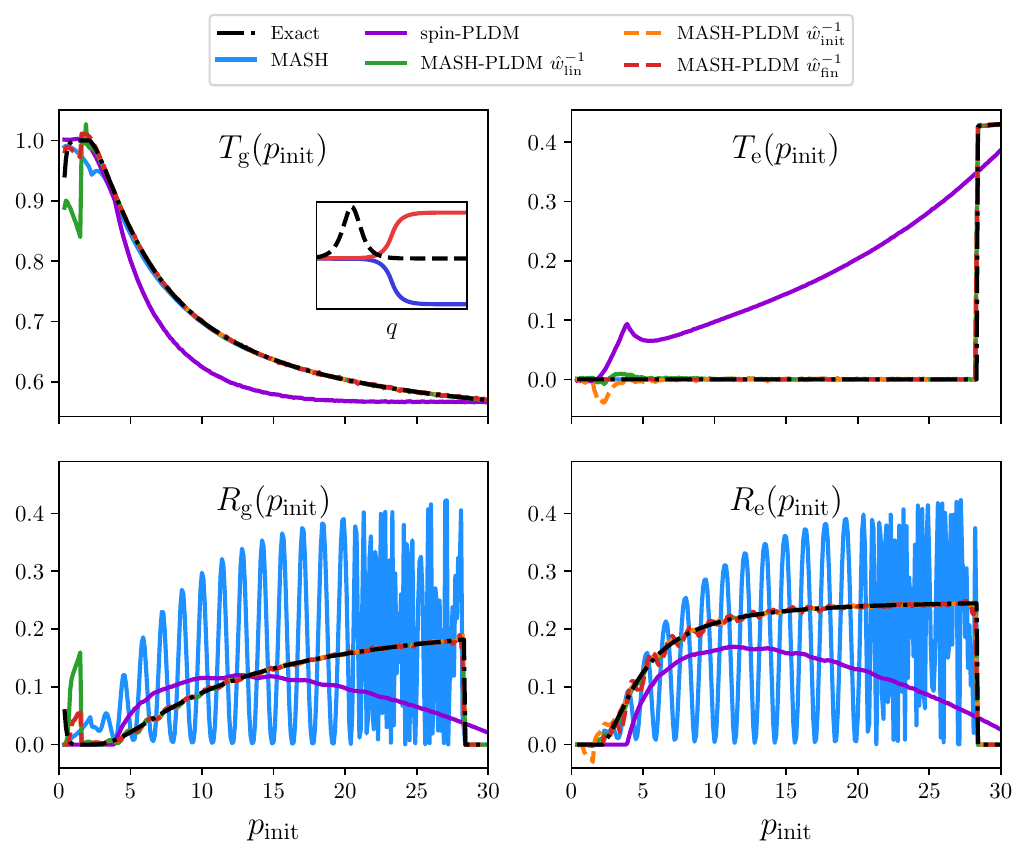}
    \caption{The scattering probabilities associated with reflection ($R$) and transmission ($T$) on each adiabatic state as a function of the initial momentum, $p_\text{init}$, for Tully’s model III\@. The inset depicts the adiabatic potentials and nonadiabatic coupling vector.
    }
    \label{fig:tully3_scatter}
\end{figure}

The scattering probabilities as a function of initial momentum are shown in Fig.~\ref{fig:tully3_scatter}.
As in Sec.~\ref{sec:Tully2}, we used energy-based initial conditions for MASH-PLDM with Sobol sampling. %

Again, the MASH-PLDM results using the final inverse kernel are the most accurate.
Not only do they correct the errors of spin-PLDM, which like SQC,\cite{Cotton2013mapping} and other mean-field mapping approaches cannot capture the step-like nature of the quantum result,
but they far surpass the MASH results, which like FSSH, show strong oscillations.\cite{MASH,Tully1990hopping,Subotnik2011decoherence}
This problem is traditionally associated with the concept of overcoherence
and a variety of decoherence corrections have been proposed to improve the result.\cite{Granucci2007FSSH,Subotnik2011AFSSH,MASH,Runeson2025decoherence}
It is revealing to observe that MASH-PLDM can capture the correct behavior without requiring \emph{ad hoc} decoherence corrections.

\begin{figure}
    \centering
    \includegraphics[width=0.7\linewidth]{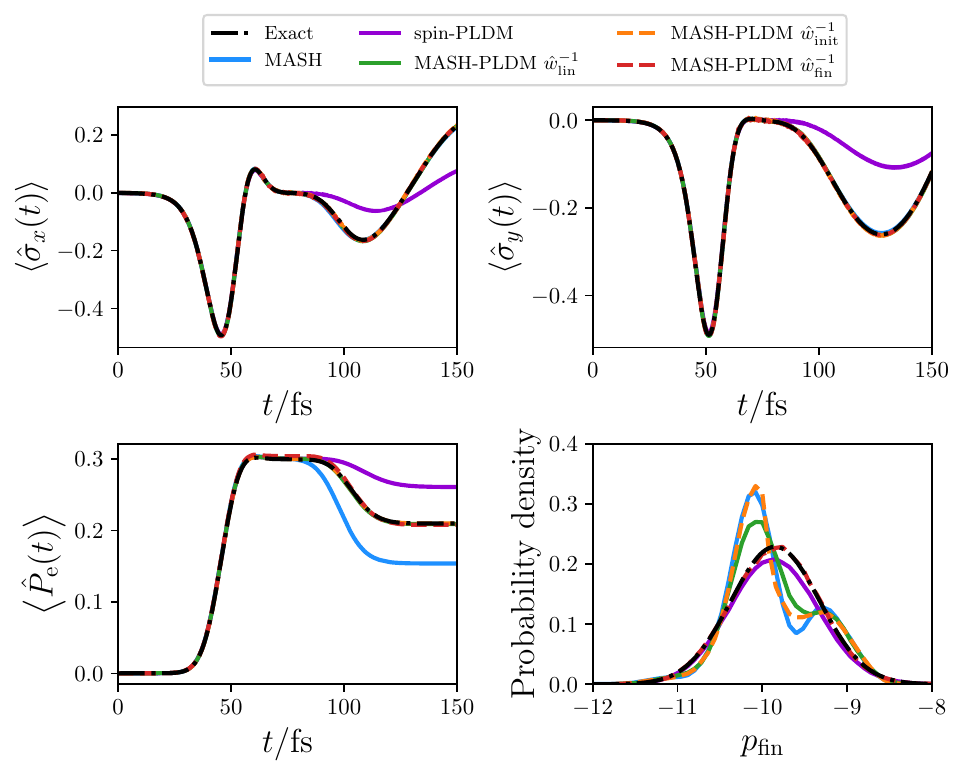}
    \caption{Time-dependent electronic observables and the final momentum distribution in Tully's model III\@.
    Initial wavepackets are given by Eq.~\eqref{eq:nucl_wavepacket_tully} with $p_\mathrm{init} = 10$.
    }
    \label{fig:tully3_pop_coh}
\end{figure}

Figure~\ref{fig:tully3_pop_coh} presents time-resolved results for a simulation of wavepacket evolution.
In particular, it shows that like original MASH (but unlike FSSH),\cite{MASH} MASH-PLDM is able to capture the electronic coherences generated by the double passage through the coupling region with remarkable accuracy.
However, whereas original MASH (like FSSH) did not capture the correct populations (unless decoherence corrections were applied),
MASH-PLDM obtains a perfect result.
This good behavior is observed for all choices of inverse kernels.
However, it is again seen that using the inverse kernel with a projection at the final time is preferable when describing the final momentum distribution of the reflected component with energy-based initial conditions.

\subsection{Initialization in a coherence}
\label{sec:coherence}

One of the important differences between MASH-PLDM and the original MASH method is its treatment of trajectories while the system is in an electronic coherence.
To highlight these effects,
we study a problem introduced by Mannouch and Kelly in \Refx{Mannouch2024coherence}.

The Hamiltonian %
is defined as in \eqn{Hdia} with diabatic potentials and couplings
\begin{subequations}\begin{align}
    V_{11}(q) &= \tfrac{1}{2}kq^2 - (kq_0q - \tfrac{1}{2}\varepsilon) , \\
    V_{22}(q) &= \tfrac{1}{2}kq^2 + (kq_0q - \tfrac{1}{2}\varepsilon) , \\
    V_{12}(q) &= b\,\exp\sbk{-a\left(q-\tfrac{\varepsilon}{2kq_0}\right)} ,
\end{align}\end{subequations}
where $k=0.02$, $b=0.01$, $a=3$, $q_0=-2$, $\varepsilon=0.01$, and $m=\num{20000}$ (all in atomic units).
The dynamics are initialized in a factorized state corresponding to the electronic density matrix (given in the diabatic representation)
\begin{equation}
    \hat{A} = \begin{pmatrix}
        0.8 & 0.4 \\
        0.4 & 0.2
    \end{pmatrix}
\end{equation}
and the vibrational ground state of the diabatic potential $V_{11}(q)$.
In addition to the time-dependent populations, various coherence estimators are used, as defined in \Refx{Mannouch2024coherence}:
\begin{subequations}
\begin{align}
    |\rho_\mathrm{ge}(t)| &= \frac{1}{2}\sqrt{\braket{\hat{\sigma}_x(t)}^2 + \braket{\hat{\sigma}_y(t)}^2} \\
    \mathcal{C}_\mathrm{ge}(q, t) &= \frac{1}{2}\sqrt{\braket{\hat{\sigma}^{(q)}_x(t)}^2 + \braket{\hat{\sigma}^{(q)}_y(t)}^2} \label{Cge_qt} \\
    \mathcal{C}_\mathrm{ge}(t) &= \int \d q_t\, \mathcal{C}_\mathrm{ge}(q_t, t),
\end{align}
\end{subequations}
where $\hat{\sigma}^{(q)}_j = \ketbra{q}{q}\otimes\hat{\sigma}_j(q)$ is a product of an adiabatic Pauli spin matrix and a projector onto the nuclear position state $\ket{q}$.

\begin{figure}
    \centering
     \includegraphics[width=0.45\linewidth]{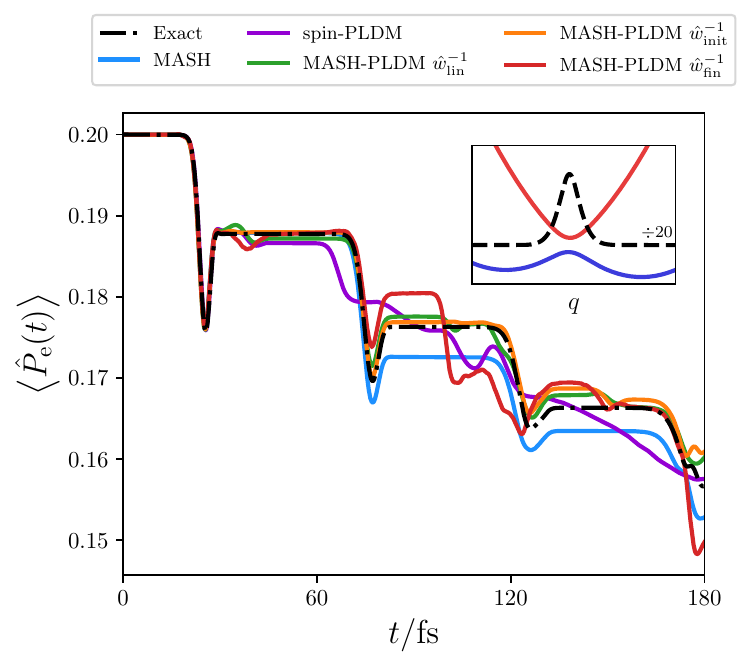}
    \caption{Time-dependent population of the upper state in the model defined in Sec.~\ref{sec:coherence}.
    The inset shows the adiabatic potentials and nonadiabatic coupling.
    }
    \label{fig:JM_populations}
\end{figure}

Figure~\ref{fig:JM_populations} shows the time evolution of the excited-state population computed by various methods. %
Whereas MASH and MASH-PLDM capture the step-like nature of the result, spin-PLDM (with focused initial conditions)\cite{spinPLDM2} smears the time-dependent populations. 
However, whereas MASH-PLDM with the initial inverse kernel offers an improvement in accuracy over the original MASH method,
results with the final inverse kernel have steps {at about 40 and 100\,fs, corresponding to trajectories evolving on the average surface reaching the avoided crossing at unphysical times.} %
Overall, therefore, combining the results of this section with those of Sec.~\ref{sec:scattering_models},
we find that one should use the initial inverse kernel in all cases where time-dependent properties are computed.
However, in scattering problems, where the result of interest is defined in terms of a long-time limit, it is even better to use the final inverse kernel along with energy-based initial conditions.

\begin{figure}
    \centering
     \includegraphics[width=0.45\linewidth]{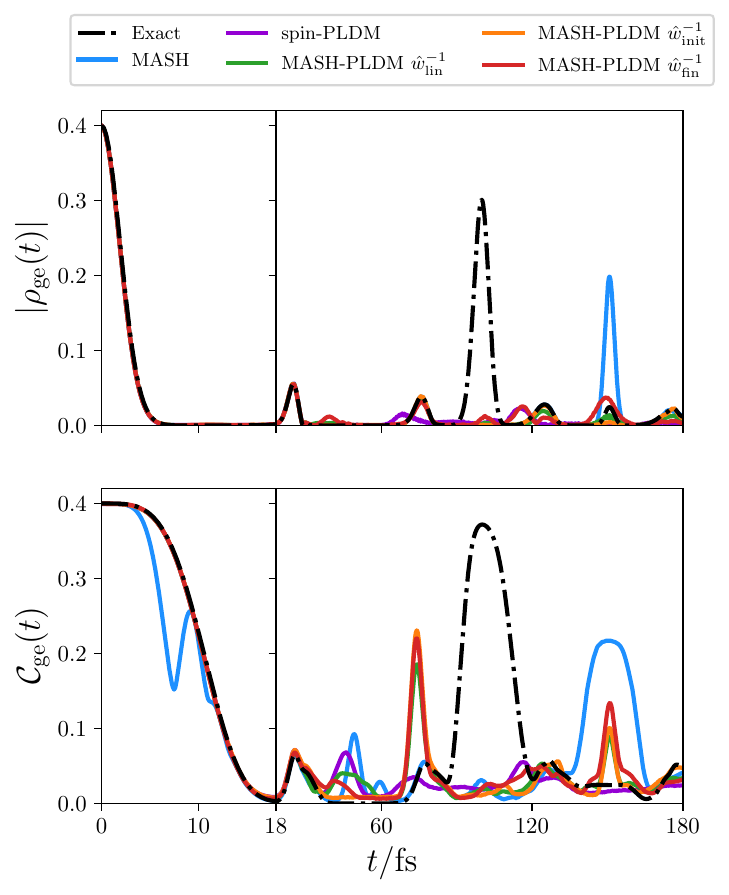}
    \caption{Evolution of an initial electronic coherence in the model of Sec.~\ref{sec:coherence}.}
    \label{fig:JM_coherence}
\end{figure}

The results for the time-dependence of coherence estimators are even more revealing.
As \Refx{Mannouch2024coherence} pointed out, MASH (like FSSH and other linearized methods) fails to describe the correct short-time decay of the coherence $\mathcal{C}_\text{ge}(t)$.
However, like spin-PLDM, MASH-PLDM captures the short-time behavior correctly of the initial decoherence as well as the peak at 20\,fs.
None of the trajectory-based methods is able to describe the %
recoherence at about 100\,fs, which would require nuclear interference effects
{beyond the reach of methods based on the QCLE.}

\begin{figure}
    \centering
    \includegraphics[width=\linewidth]{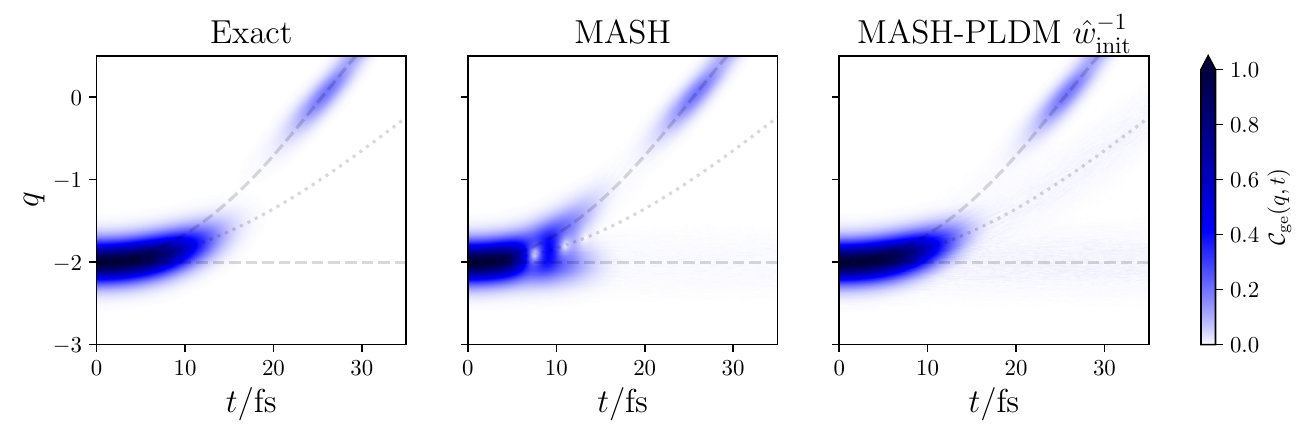}
    \caption{Spatial contributions to coherence [\eqn{Cge_qt}].
    Gray lines indicate the progress of a trajectory on a single adiabat (dashed) or the averaged surface (dotted).
    }
    \label{fig:JM_spatial_coh}
\end{figure}

To demonstrate why MASH-PLDM is able to correctly describe the initial decoherence,
Fig.~\ref{fig:JM_spatial_coh} shows the time-evolution of the trajectories which contribute to the coherence signal.
The distribution is much more accurately captured by MASH-PLDM than by MASH due to the trajectories which move on the averaged surface.
In addition, there are contributions to the coherence beginning around 20\,fs, which are induced by trajectories on the upper surface passing through the avoided crossing. %

This positive result demonstrates that our formulation of MASH-PLDM accomplishes the key requirements for a correct description of initial electronic coherence identified in \Refx{Mannouch2024coherence}.
In particular, it justifies that evolving on an average surface can be a benefit for trajectory-based nonadiabatic dynamics.

\subsection{Spin--boson models} \label{sec:spinboson}

Finally, we apply our methods to a set of multidimensional spin--boson models 
employed in previous work. \cite{Cotton2016SBmodels,linearized,spinmap,spinPLDM1}
The Hamiltonian in the diabatic basis is 
\begin{subequations}\begin{align}
    \hat{H} &= \varepsilon\hat{\sigma}_z^\text{dia} + \Delta\hat{\sigma}_x^\text{dia} + \sum_{j=1}^{f}\left[ \frac{1}{2}(\hat{p}^2_{j} + \omega_{j}^2 \hat{q}^2_{j}) + c_{j}\hat{q}_{j}\hat{\sigma}_z^\text{dia}\right] ,
\end{align}\end{subequations}
where $\varepsilon$ is (twice) the energy bias, $\Delta$ is the diabatic coupling, and $\omega_{j}$ and $c_{j}$ are the frequency and the coupling strength of the $j$th mode of the bath with $f$ modes.
Here we employ the diabatic Pauli matrices $\pauli_\mu^\text{dia}$, which obey the same properties as the adiabatic matrices at a fixed nuclear configuration.
The spectral density determines the distribution of the frequencies and coupling strength. Here, we use an Ohmic spectral density, 
$J(\omega)=\tfrac{\pi}{2}\xi\omega\,\eu{-\omega/\omega_\mathrm{c}}$,
where $\omega_\mathrm{c}$ is the characteristic frequency and $\xi$ is the Kondo parameter. 
We discretize the continuous bath into $f=100$ or 400 modes, following the discretization scheme %
of Ref.~\onlinecite{RPMDrate}, %
and initialize the bath modes from the thermal Wigner distribution
\begin{equation}
    \rho_\mathrm{n}(\bm{q},\bm{p}) = \prod_{j=1}^{f} \alpha_j \exp\left[-\frac{\alpha_j}{\omega_j} \left(\frac{1}{2}p_{j}^2 + \frac{1}{2}\omega_{j}^2 q_{j}^2\right)\right],
\end{equation}
where $\alpha_j = 2\tanh(\tfrac{1}{2}\beta\omega_j)$ and $\beta = 1/k_\mathrm{B}T$ is the inverse temperature. Numerically exact results for the real-time quantum correlation functions were taken from \Refx{spinPLDM1}, which obtained them using the quasiadiabatic path-integral (QUAPI) technique.\cite{Makri1995QUAPI}
Note that MASH and MASH-PLDM trajectories were evolved in the adiabatic basis and the diabatic observables were computed as linear combinations of adiabatic Pauli matrices.
On the other hand, spin-PLDM (here with full-sphere initial conditions) gives equivalent results whether it is implemented in the diabatic basis or using adiabatic states with the kinematic momentum.

\begin{figure}[h]
    \centering
    \includegraphics[width=0.97\linewidth]{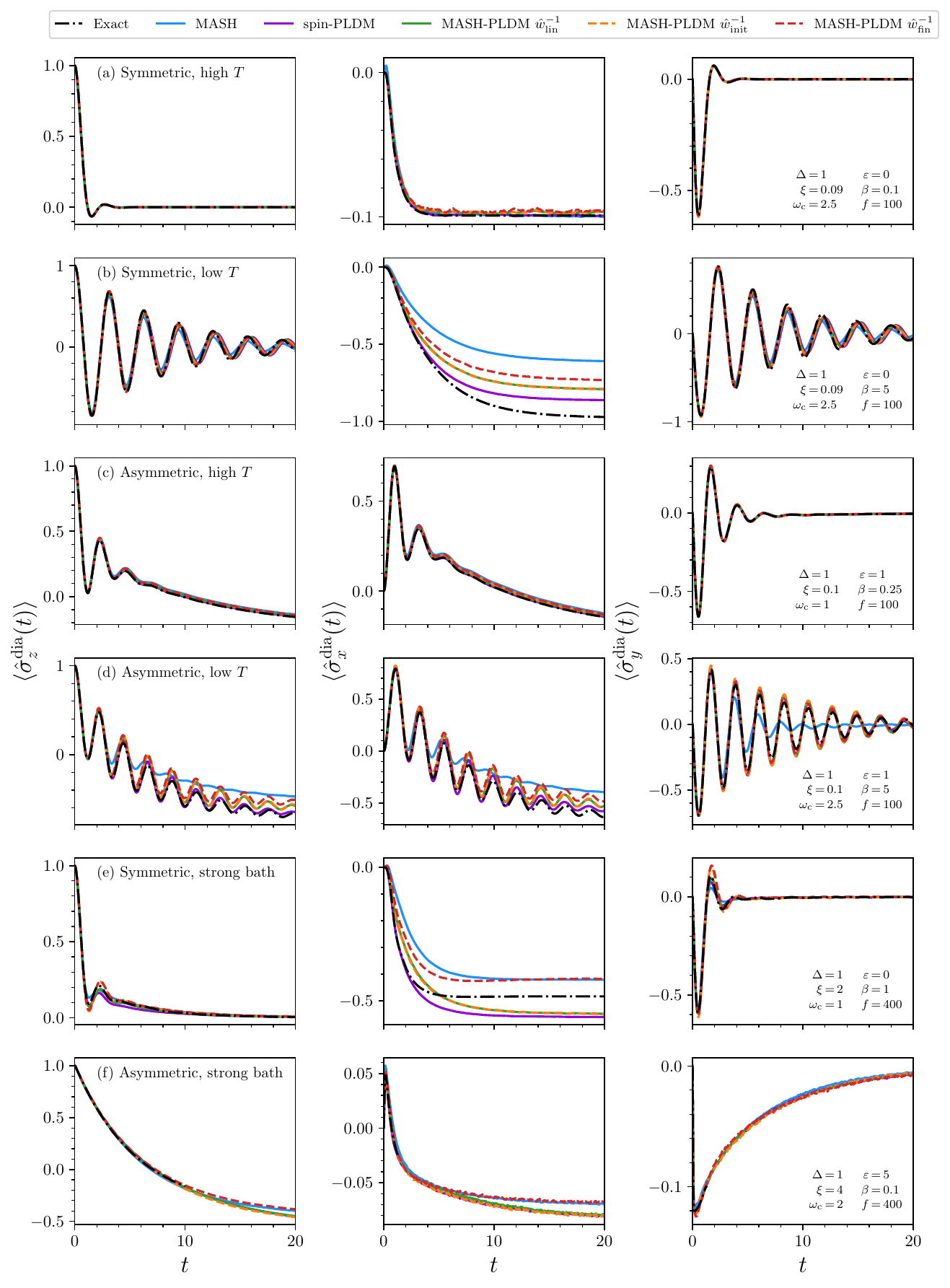}
    \caption{Time-dependent diabatic expectation values for a set of spin--boson models compared with exact QUAPI benchmark data (which is only plotted where it was deemed to be fully converged).
    The parameters of each model are given %
    in the right column.
    }
    \label{fig:spin_boson_results}
\end{figure}

Figure \ref{fig:spin_boson_results} shows the correlation functions starting in a diabatic population $\hat{A}=\half(\Id+\pauli_z^\text{dia})$ with the measurement operator %
$\hat{B}=\pauli_\mu^\text{dia}$
for six different parameter regimes, including both symmetric and asymmetric systems, low- and high-temperature limits, and strong and weak system–bath coupling.
In these problems, the spin-PLDM method is found to be remarkably accurate.\cite{spinPLDM1}
However, although its mean-field approximation does not lead to significant errors in this simple harmonic model, we have shown that it cannot be relied upon in general in Secs.~\ref{sec:scattering_models} and \ref{sec:coherence}.

In many cases, especially at high temperatures, MASH-PLDM gives results almost as accurate as spin-PLDM (regardless of the choice of inverse kernel), just as for the original MASH method.
The most difficult problems appear to be %
the two systems at low temperature, %
where MASH exhibits overdamped oscillations.
In these cases, MASH-PLDM seems to give more reliable results and in particular captures the amplitude of oscillations with a much higher accuracy.

In cases where MASH-PLDM is accurate, it typically gives consistent results with all three choices of inverse kernel.
In a couple of cases, where there are discrepancies between the linear/initial kernels and the final kernel, %
it seems that the linear/initial kernels are in best agreement with spin-PLDM\@.
This is in line with our conclusions of Sec.~\ref{sec:coherence}, in which 
we would assume that the initial kernel is the preferred choice for this case.

\section{Multi-time correlation functions for nonlinear spectroscopy}

Time-resolved spectra can be obtained by applying Fourier transforms to nonlinear response functions.\cite{MukamelBook}
In particular, many important types of spectroscopy are based on the third-order response, which can be defined 
in terms of multi-time correlation functions such as
\begin{align}
    R(t_1,t_2,t_3) = \Tr\left[ \hat{\rho}_\mathrm{n} \hat{A} \, \e^{i\hat H (t_1+t_2+t_3)} \hat{B} \, \e^{-i\hat H t_3} \, \hat{C} \, \e^{-i\hat H t_2} \hat{D} \, \eu{-i\hat{H}t_1} \right].
\end{align}

\begin{figure}
    \centering
    \tikzset{
  midarrow/.style={postaction={decorate},
    decoration={markings,
    mark=at position .25 with {\arrow[scale=1.5]{>}},
    mark=at position .75 with {\arrow[scale=1.5]{>}}
    }},
}

\begin{tikzpicture}[>=stealth]

\draw[Blue,thick,wavy,->] (-4,0) -- (-3,-1)
  node[midway,below left] {$\hat \rho_\text{n} \otimes \hat A$};
  
\draw[orange,thick,midarrow]
  (5,4) .. controls ++(-3,-0.5) and ++(0,3) .. (-4,0)
  node[pos=1, left] {$\bm{S}'(0)$}
  node[pos=0.15, above=0.3] {$\qquad\qquad \bm{S}'(t_1 + t_2 + t_3)$};

\draw[purple,thick,midarrow]
  (-3,-1) .. controls ++(0.7,0.3) and ++(-0.7,0.3) .. (-1,-1)
  node[pos=0, below] {$\quad\bm{S}(0)$};
\draw[purple,thick,midarrow]
  (0,0) .. controls ++(0.7,-0.3) and ++(-0.7,0.3) .. (3,0);
\draw[purple,thick,midarrow]
  (4,1) .. controls ++(0,1) and ++(-1,-1.5) .. (6,3)
  node[pos=1, right] {$\bm{S}(t_1 + t_2 + t_3)$};

\draw[SkyBlue,thick,-{stealth[length=3mm]}, wavy] (-1,-1) -- (0,0)
  node[midway,above left] {$\hat D$};

\node[right] at (-0.5,-0.7) {{\color{purple}$\bm{S}(t_1)$}};
  
\draw[green!50!black,thick,-{stealth[length=3mm]}, wavy] (6,3) -- (5,4)
  node[midway,above right] {$\hat B$};

\draw[green!85!black,thick,-{stealth[length=3mm]}, wavy] (3,0) -- (4,1)
  node[midway,above left] {$\ \ \hat C$};

\node[right] at (3.5,0.3) {{\color{purple}$\bm{S}(t_1 + t_2)$}};
  
\draw[black,thick,bwfwarrows]
  (-3.5,-0.5) .. controls ++(2,3) and ++(-3,-2) .. (5.5,3.5)
  node[pos=0.5, below right] {${\Bq}(t)$};

\node[right] at (-4,-2.5) {$
    R(t_1, t_2, t_3) = \Tr\!\left[ 
    {\color{Blue}\hat{\rho}_\mathrm{n} \hat{A}} \, 
    {\color{Orange}\e^{i\hat H (t_1+t_2+t_3)}}\, 
    {\color{green!50!black}\hat{B}} \, 
    {\color{purple}\e^{-i\hat H t_3}} \,
    {\color{green!85!black}\hat{C}} \, 
    {\color{purple}\e^{-i\hat H t_2}} \,
    {\color{SkyBlue}\hat{D}} \, 
    {\color{purple}\eu{-i\hat{H}t_1}}\right]$};

\end{tikzpicture}
    \caption{An example of a multi-time correlation function that %
    contributes to third-order nonlinear response theory.
    To evaluate the expression using MASH-PLDM, two spin vectors are propagated which experience different interactions along the forward and backward paths.
    }
    \label{fig:nonlinear}
\end{figure}

Unlike the original MASH method and other linearized approaches, which are based on mapping the observables (for which at most two are allowed as explained in Appendix~\ref{app:MASHLSC}),
it is easy to extend the derivation of MASH-PLDM to treat multi-time correlation functions.
As before,
the nuclear degrees of freedom are linearized to give one set of $\Bq$ and $\Bp$ variables
and
by introducing an inverse mapping kernel for each of the forward and backward paths, we obtain two separate spin vectors, as shown in Fig.~\ref{fig:nonlinear}.
Finally, one evaluates the electronic trace over the propagators and multiple observable operators at the appropriate time points.
We will illustrate this approach with a specific example.

Inspired by the seminal experiments of Zewail,\cite{NaI2,Zewail2000nobel}
we will consider a pump--probe signal for a model of the NaI molecule with 3 states: the ionic and covalent states (corresponding to the ground and first-excited states), which are coupled leading to an avoided crossing, and an uncoupled high-energy final state.
The diabatic potentials and couplings are defined as\cite{NaI-pot,NaI2}
\begin{subequations}\begin{align}
    V_{11}(q) &= A_1 \, \eu{-\beta_1(q - q_0)}, \\
    V_{22}(q) &= \left[A_2 + \left(\frac{B_2}{q}\right)^8\right]\eu{-q/\rho} - \frac{e^2}{q} - \frac{e^2(\lambda_+ + \lambda_-)}{2q^4} - \frac{C_2}{q^6} - \frac{2e^2\lambda_+\lambda_-}{q^7} + \Delta E_0, \\
    V_{12}(q) &= A_{12}\,\eu{-\beta_{12}(q-q_x)^2}, \\
    V_{\rm f}(q) &= D_e\left[1 - \eu{-\alpha(q-q_e)}\right]^2 - D_e + E_{\rm exc},
\end{align}\label{eq:NaI_potential}\end{subequations}
where $A_1 = 0.813\,\text{eV}$, $A_2 = 2760\,\text{eV}$, $A_{12} = 0.055\,\text{eV}$, $B_2 = 2.398\,\text{eV}^{1/8}\text{\AA}$, $C_2 = 11.3\,\text{eV}\text{\AA}^6$, $\beta_1 = 4.08\,\text{\AA}^{-1}$, $\beta_{12} = 0.6931\,\text{\AA}^{-2}$, $\lambda_+ = 0.408\,\text{\AA}^3$, $\lambda_{-} = 6.431\,\text{\AA}^3$, $\Delta E_0=2.075\,\text{eV}$, $\rho=0.3489\,\text{\AA}$, $D_e=1863\,\text{cm}^{-1}$, $\alpha = 0.83\,\text{\AA}^{-1}$, $q_0=2.67\,\text{\AA}$, $q_e = 4\,\text{\AA}$, $q_x=6.93\,\text{\AA}$, $e$ is the electron charge,
and $E_{\rm exc} = \SI{17000}{cm^{-1}}$ is the excitation energy from the dissociated covalent state. %
The adiabatic states $\ket{\Phi_{\rm g}}$, $\ket{\Phi_{\rm e}}$ and $\ket{\Phi_{\rm f}}$ are obtained by diagonalizing the diabatic potential energy matrix.

In the limit that the pump pulse is instantaneous, we can set $t_1=0$ and the relevant multi-time correlation function
for modeling the excited-state absorption becomes
\begin{align}  \label{ESA}
    R(0,t_2,t_3) &= \Tr\left[ \hat{\rho}_\mathrm{n} \, \hat\mu_\mathrm{ge} \, \e^{i\hat H (t_2+t_3)} [\hat\mu_\mathrm{gf} + \hat\mu_\mathrm{ef}] \, \e^{-i\hat H t_3} \, [\hat\mu_\mathrm{fg} + \hat\mu_\mathrm{fe}] \, \e^{-i\hat H t_2} \, \hat\mu_\mathrm{eg} \right] ,
\end{align}
where $\hat{\rho}_\mathrm{n}=\ketbra{\chi_0}{\chi_0}$ is the initial density matrix
defined by %
the ground-state vibrational wavefunction, $\ket{\chi_0}$. %
The excitation operators are $\hat\mu_{\ell\ell'}=\ketbra{\Phi_\ell}{\Phi_{\ell'}}$ for $\ell,\ell'\in\{\mathrm{g,e,f}\}$, which for simplicity we assume to be independent of the nuclear coordinate (known as the Condon approximation).
The pump--probe spectrum can then be calculated as\cite{MukamelBook}
\begin{equation}
    I_\mathrm{PP}(\omega_3,t_2)=-\frac{1}{2\pi}{\Im}\left[\int_0^\infty\d t_3\, \frac{2\Im[R(0,t_2,t_3)]}{R(0,t_2, 0)} \,\eu{i\omega_3 t_3} \right].
\end{equation}

The response function \eqn{ESA}
can be calculated within the MASH-PLDM formalism developed in this work. %
For instance, using the inverse kernel with projections at initial times,
\begin{subequations}
\begin{multline}
    R^\text{init}(0,t_2,t_3) = \sum_{\ell,\ell'\in\{\mathrm{g},\mathrm{e}\}} \int \frac{\d\Bq\,\d\Bp}{(2\pi)^f}\,\d\bm{S}\,\d\bm{S}' \rho_\mathrm{n}(\Bq,\Bp)
    \\ \times \braket{\Phi_\mathrm{e} | \winv_{\Bq,{\rm init}}(\bm{S}')^\dagger \hat{U}(t_2+t_3,0)^\dagger \hat{\mu}_{\ell'\mathrm{f}} \hat{U}(t_2+t_3,t_2) \hat{\mu}_{\mathrm{f}\ell} \hat{U}(t_2,0) \winv_{\Bq, {\rm init}}(\bm{S}) | \Phi_\mathrm{e}},
    \label{Rinit}
\end{multline}
or using projections at final times,
{\begin{multline}
    R^\text{fin}(0,t_2,t_3) = \sum_{\ell,\ell'\in\{\mathrm{g},\mathrm{e}\}} \int \frac{\d\Bq\,\d\Bp}{(2\pi)^f}\,\d\bm{S}\,\d\bm{S}' \rho_\mathrm{n}(\Bq,\Bp)
    \\ \times \braket{\Phi_\mathrm{e} | \hat{U}(t_2+t_3,0)^\dagger \winv_{\Bq,{\rm fin}}(\bm{S}'(t_2+t_3))^\dagger  \hat{\mu}_{\ell'\mathrm{f}} \hat{U}(t_2+t_3,t_2) \hat{\mu}_{\mathrm{f}\ell}\, \winv_{\Bq, {\rm fin}}(\bm{S}(t_2)) \hat{U}(t_2,0)  | \Phi_\mathrm{e}}.
\end{multline}}
\end{subequations}
{The case with the linear inverse kernel is equivalent to \eqn{Rinit} except with $\winv_\text{lin}$ substituted for $\winv_\text{init}$.}
As before, the nuclear density operator %
is Wigner transformed to give $\rho_\mathrm{n}(\Bq,\Bp)$. In this work, we additionally employ a harmonic approximation around the minimum of the ground-state potential to obtain a simple Gaussian form. %
MASH-PLDM trajectories are run within the manifold of states spanned by $\mathrm{g}$ and $\mathrm{e}$ during the $t_2$ time and then propagated for $t_3$ using
$\pder{\Bp}{t}=\half[-\grad V_{\mathrm{f}}(\Bq) + \bm{F}(\Bq,\bm{S}')]$
as the nuclear force
[where $\bm{F}(\Bq,\bm{S}')$ is the MASH force %
in the $\{\mathrm{g},\mathrm{e}\}$ subspace]
with only one set of mapping variables.
In practice, we run an ensemble of trajectories for the maximum value of $t_2$ required and spawn short $t_3$ trajectories at a range of intermediate $t_2$ times.
The propagators are defined as %
$\hat{U}(\ell\epsilon,k\epsilon) = \eu{-i\epsilon\hat V(\Bq_\ell)/2}\,\eu{-i\epsilon\hat V(\Bq_{\ell-1})}\cdots\eu{-i\epsilon\hat V(\Bq_{k+1})}\,\eu{-i\epsilon\hat V(\Bq_k)/2}$, using the full 3-state potential.
The inverse kernels, %
which are defined by Eq.~\eqref{eq:variants} in the $\{\rm g, e\}$ subspace,
can be padded with zeros to account for the uncoupled $\rm f$ state %
in order to evaluate the expression using matrix algebra.

For comparison, we generated an exact quantum-mechanical benchmark by calculating $R(0,t_2,t_3)$ using the split-operator method, evolving an initial Gaussian wavepacket corresponding to the harmonic approximation of the ground state wavefunction.

In practice, as we only run the trajectory and split-operator simulations for a finite $t_3$, we multiply the response functions by an exponential damping function, $\eu{-\gamma t_3}$, before Fourier transforming in order to obtain smooth spectra. %
The pump--probe spectra shown in this section %
are evaluated with $\gamma=0.003$.

\begin{figure}
    \centering
    \includegraphics[width=\linewidth]{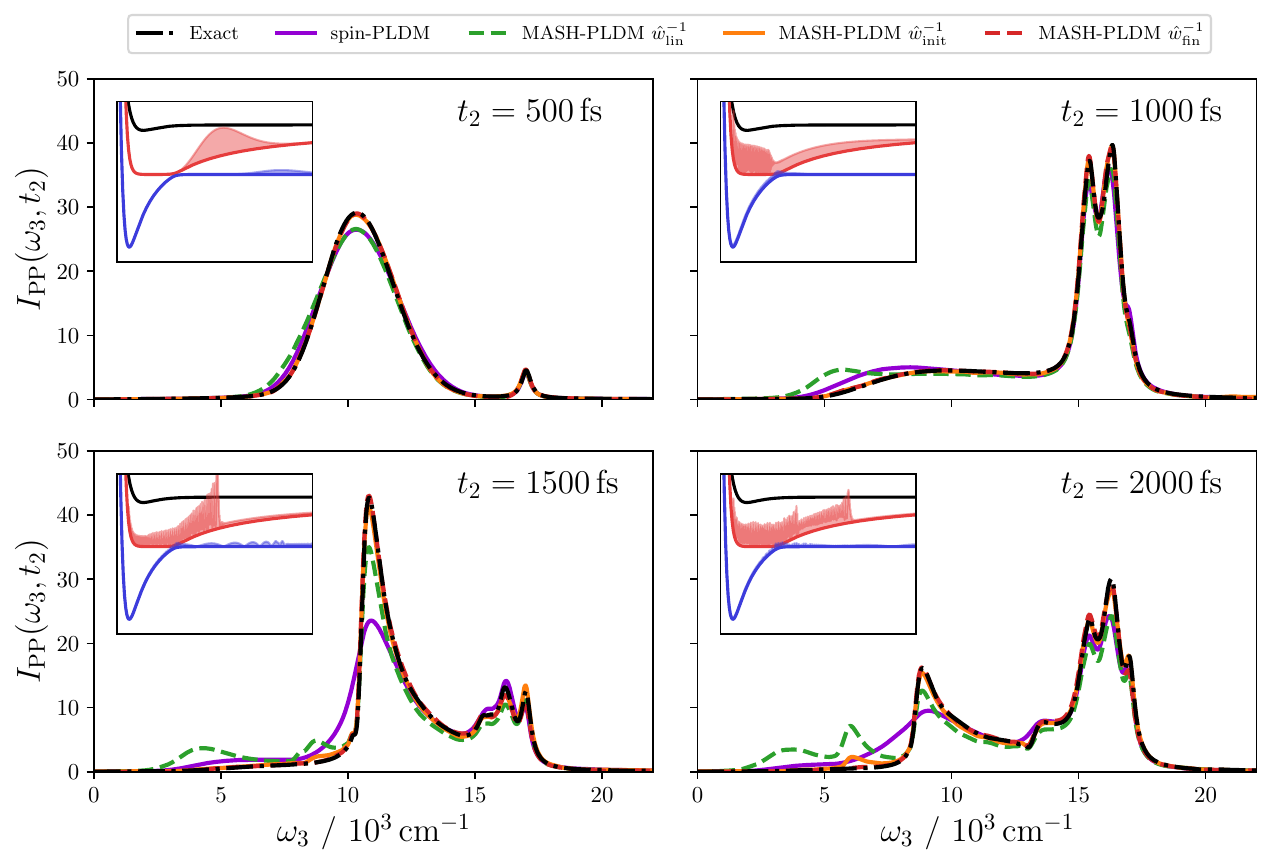}
    \caption{Pump--probe spectra calculated for the NaI model at four values of $t_2$.
    The insets depict the adiabatic potentials and the position of the wavepackets (absolute value) according to the exact quantum dynamics. 
     }
    \label{fig:NaI}
\end{figure}

\begin{figure}
    \centering
    \includegraphics[width=\linewidth]{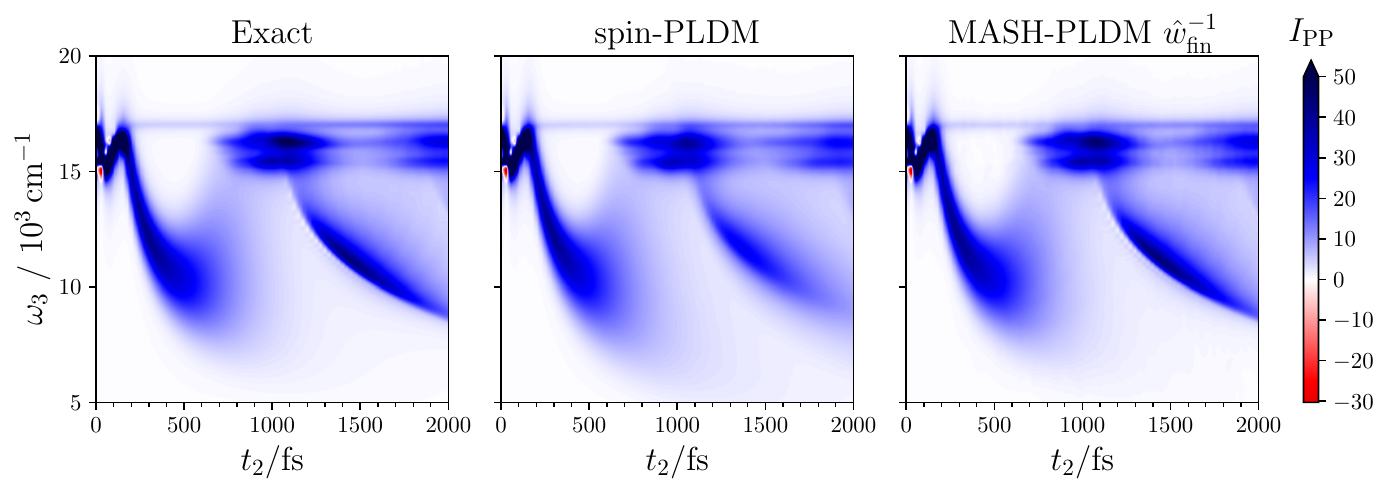}
    \caption{Time-evolution of the pump--probe spectrum calculated for the NaI model. %
    }
    \label{fig:NaI_3d}
\end{figure}

The results of MASH-PLDM are compared to the quantum-mechanical benchmark in Fig.~\ref{fig:NaI}.
The probed spectrum is presented after (roughly) one, two, three, and four transitions of the nonadiabatic coupling region, as indicated by the insets which show the location of the wavepacket from the quantum simulation.
In particular, {at $t_2=500\,\mathrm{fs}$,} one can see that the wavepacket has split onto the two states resulting in a bimodal spectrum, which is well captured by the MASH-PLDM method, especially when using the inverse kernel with projections at {either the initial or final times}.
{At later times, the spectrum becomes more complicated as the wavepacket spreads,}
{and it appears that the projection at final time is most accurate as it avoids a spurious peak at about $\omega_3\approx6\times10^3\,\mathrm{cm}^{-1}$ observed in the results at $t_2=2000\,\mathrm{fs}$ with $\winv_\text{init}$, which comes from trajectories moving on the average surface and which are eliminated by projecting at the final time.}
{The reason why the final projection is so accurate here is because,
in this problem, trajectories initialized with at least one spin on the lower hemisphere are trapped in the potential well and never reach the nonadiabatic coupling region. %
In this way, even without an initial projection, only trajectories initialized on the upper surface contribute to the correlation function, meaning that one can benefit further 
from the consistency imposed by the final projection.}
{In Fig.~\ref{fig:NaI_3d}, we construct a time-series of data showing how the spectral peak splits, reforms, and splits again as the system crosses and recrosses the nonadiabatic coupling region.
The results of MASH-PLDM with the final projection are practically indistinguishable from the exact quantum benchmark.}

In contrast, under spin-PLDM dynamics, the spectra in Fig.~\ref{fig:NaI} are artificially broadened.
This is because, like all mean-field dynamical methods including Ehrenfest, the original PLDM, and FBTS,
the trajectories fail to correctly describe nuclear wavepacket branching after passing through a nonadiabatic coupling region.
This problem is already visible after the first crossing, but becomes more significant after multiple crossings.
A similar problem affects MASH-PLDM with the linear kernel, as was already discussed in Sec.~\ref{sec:Tully1}.
Note that the %
high-frequency peaks are not adversely affected, %
as the ground-state potential is flat in the asymptotic region, making the excitation frequency insensitive to the nuclear distribution.
Although the spin-PLDM results still capture the time-evolved pump--probe spectrum in Fig.~\ref{fig:NaI_3d} reasonably well,
the results are noticeably smeared and we assume that the problems will become even more severe in more realistic multidimensional molecular systems.

It was to avoid these problems of mean-field dynamics that we turned to
surface-hopping trajectories such as those of MASH-PLDM using the inverse kernels with projections.
In fact, other surface-hopping methods have been previously proposed for simulating nonlinear spectroscopy, although they only actually employed the surface-hopping algorithm during the $t_2$ delay time, whereas ground-state dynamics were used for the $t_1$ and $t_3$ times.\cite{Tempelaar2018FSSH}
In contrast, the MASH-PLDM approach can describe nonadiabatic dynamics not just during the $t_2$ evolution, but also during $t_1$ and $t_3$.
The advantages of this more powerful formulation will be further explored in following studies.

\section{Conclusions}

We have successfully combined the advantages of two nonadiabatic dynamics methods, MASH and PLDM, to create MASH-PLDM\@.
Unlike the original PLDM method\cite{Huo2011densitymatrix} and other mapping approaches,\cite{Sun1998mapping,Wang1999mapping,Stock2005nonadiabatic,Hsieh2012FBTS,Hsieh2013FBTS,Miller2016Faraday,identity,spinmap,multispin,spinPLDM1,spinPLDM2,nonlinear,Gao2020mapping} which suffered from mean-field forces, the MASH-PLDM trajectories make deterministic hops between states, leading to a correct description of wavepacket branching
without requiring expensive iterative approaches. \cite{Huo2012PLDM}
Two key differences between MASH-PLDM and the original MASH method are in their treatment of coherences.
First, unlike in the original MASH method, MASH-PLDM may propagate trajectories on an averaged surface in order to describe a coherence,  
which gives dynamics closer to the quantum truth than moving on either state alone.
Second, MASH-PLDM inherits the ability to simulate multi-time correlation functions, such as those required for nonlinear spectroscopy.

MASH-PLDM retains a rigorous connection to the QCLE (within the momentum-jump approximation)\cite{Kapral1999QCLE,Kapral2015QCL} and recovers it in the short-time limit.
Unlike other surface-hopping solutions to the QCLE (which aim to capture the QCLE behavior for all times), MASH-PLDM does not require the introduction of a stochastic jumping algorithm leading to the accumulation of Monte-Carlo weights and instabilities.\cite{Kapral2015QCL}

In fact, we have presented three different versions of MASH-PLDM in this work.
The version based on the linear inverse kernel has a simpler mathematical structure and thus we have focused on this during our derivation.
However, the two versions including projections at the initial or final time are just as easy to work with in practice (if not even easier as the initial projection allows one to sample only half the Bloch sphere).
Our results suggest that the initial projection is to be favored in most applications, both from the point of view of efficiency and accuracy.
However, in special cases, such as scattering problems from a well-defined energy, {or other cases in which the specific details of the problem implicitly select relevant trajectories with the appropriate initial conditions}, even more accurate results can be obtained from the final projection. %

While the original MASH method has a similar convergence behavior to the commonly-used FSSH approach,
MASH-PLDM requires %
more trajectories to converge to a similar level of acceptable statistical noise.
One reason for this is that there are two spin spheres to sample rather than just one.
This seems to imply that using MASH-PLDM will entail a significant computational cost relative to current nonadiabatic dynamics methods.
We note, however, that many properties of interest to MASH-PLDM will be short-time properties, such as ultrafast spectroscopy, and thus although it requires many trajectories, they will at least be short.
Perhaps its saving grace is that recent developments in machine learning for nonadiabatic dynamics\cite{Westermayr2020perspective,MLNACs,Dupuy2024MLNACs}
will make it possible to actively learn the energies, gradients and couplings in the relevant parts of space
to enable thousands of trajectories to be run at a fraction of the cost.

In this study, we have presented proof-of-principle benchmarks to test the accuracy of the MASH-PLDM method.
Although it seems to perform well for spin--boson models and simple one-dimensional scattering models,
these systems can also be treated reasonably well by other methods. %
We believe that its real power will only be fully demonstrated in complex molecular systems involving bond breaking, where unlike in simplified harmonic models of the condensed phase, one cannot get away with mean-field dynamics, and the benefits of surface-hopping trajectories will be clearly seen.
In particular, the ability of MASH-PLDM to properly describe coherences and multi-time correlation functions
makes it a particularly powerful method for the simulation of nonlinear spectroscopy in molecular systems.
We will tackle such problems in future work.

\section*{Supplementary Material}
See the supplementary material for results pertaining to the convergence behavior of the methods.

\section*{Acknowledgements}
The authors thank Jonathan Mannouch and Daniele Furlanetto for helpful discussions.

\section*{Author declarations}
\subsection*{Conflict of Interest}
The authors have no conflicts to declare.
\subsection*{Author contributions}

\textbf{Jan Obermeier:} Methodology (lead); Software (lead); Formal analysis (lead); Data Curation (equal); Investigation (equal); Visualization (lead); Writing -- original draft (supporting); Writing -- review \& editing (supporting).
\textbf{Kasra Asnaashari:} Methodology (supporting); Software (supporting); Data Curation (equal); Investigation (equal); Supervision (supporting); Writing -- original draft (supporting); Writing -- review \& editing (supporting).
\textbf{Jeremy O. Richardson:} Conceptualization (lead); Formal analysis (equal); Methodology (equal); Supervision (lead); Writing -- original draft (lead); Writing -- review \& editing (lead).

\section*{Data Availability}
The data that support the findings of this study are available within the article and its supplementary material.

\appendix

\section{Derivation of the Wigner-transformed correlation function}
\label{app:Wigner}

First, by inserting complete sets of position and momentum states, the quantum-mechanical correlation function can be written as
\begin{align}
    C_{AB}(t) %
    &= \int \d\q_0 \d\q_0' \d\p_0 \d\p_0' \d\q_N \d\q_N' \d\p_N \d\p_N'
    \tr\Big[\braket{\q_0|\hat{\rho}_\mathrm{n}\hat{A}|\q_0'}
    \braket{\q_0'|\p_0'} \braket{\p_0'|\e^{i\hat{H}t}|\q_N'} 
    \nonumber \\ &\quad\times
    \braket{\q_N'|\p_N'} \braket{\p_N'|\hat{B}|\p_N} \braket{\p_N|\q_N} 
    \braket{\q_N|\e^{-i\hat{H}t}|\p_0} \braket{\p_0|\q_0} \Big] .
\end{align}
Second, using $\braket{\q|\p} = (2\pi)^{-f/2}\,\e^{i\q\cdot\p}$,
\begin{align}
    C_{AB}(t) &= \int
    \d\Bq_0 \d\Dq_0 \frac{\d\Bp_0}{(2\pi)^f} \d\Dp_0 \d\Bq_N \d\Dq_N \frac{\d\Bp_N}{(2\pi)^f} \d\Dp_N
    \tr\Big[ \braket{\Bq_0+\thalf\Dq_0|\hat{\rho}_\mathrm{n}\hat{A}|\Bq_0-\thalf\Dq_0}
    \nonumber \\ &\quad\times
    \eu{i\q_0'\cdot\p_0'} \braket{\p_0'|\e^{i\hat{H}t}|\q_N'} \eu{i\q_N'\cdot\p_N'} \braket{\Bp_N-\thalf\Dp_N|\hat{B}|\Bp_N+\thalf\Dp_N} 
    \eu{-i\p_N\cdot\q_N} \braket{\q_N|\e^{-i\hat{H}t}|\p_0} \eu{-i\p_0\cdot\q_0} \Big] .
\end{align}
Third, using $\p\cdot\q - \p'\cdot\q' %
= \Bp\cdot\Dq + \Dp\cdot\Bq$,
\begin{align}
    C_{AB}(t) %
    &= \int
    \d\Bq_0 \frac{\d\Bp_0}{(2\pi)^f} \d\Dp_0 \d\Bq_N \d\Dq_N \frac{\d\Bp_N}{(2\pi)^f}
    \tr\Big[ [\hat{\rho}_\mathrm{n}\hat{A}]_\mathrm{W}(\Bq_0,\Bp_0)
    \,\eu{-i\Bq_0\cdot\Dp_0} \braket{\p_0'|\e^{i\hat{H}t}|\q_N'}
    \nonumber \\ &\quad\times
    \hat{B}_\mathrm{W}(\Bq_N,\Bp_N) \,\eu{-i\Bp_N\cdot\Dq_N} 
    \braket{\q_N|\e^{-i\hat{H}t}|\p_0} \Big] ,
\end{align}
where subscript `W' indicates a partial Wigner transform [\eqn{eq:wigner} or the equivalent with momentum and position variables interchanged].
Finally, inserting further identities into the propagators, making the linearization approximation and collecting the phases recovers \eqn{eq:QCLE}.

\section{Derivation of PLDM} \label{app:perturbation}

Introducing the short-hand notation $\hat{V}_k=\hat{V}(\Bq_k)$ and $\hat{W}_k=\half\nabla\hat{V}(\Bq_k)\cdot\Dq_k$ with $\nabla\hat{V}$ defined by Eq.~\eqref{eq:force},
we can write the propagator from Eq.~\eqref{eq:T} as
\begin{align} \label{eq:Tagain}
    \hat{T} &= \eu{-i\epsilon\hat{W}_{N}} \eu{-i\epsilon\hat{V}_{N}} \cdots \eu{-i\epsilon\hat{W}_1} \eu{-i\epsilon\hat{V}_1} .
\end{align}
Approximating this expression within the framework of time-dependent perturbation theory, in which we treat $\hat{V}_k$ as the reference Hamiltonian and $\hat{W}_k$ as the perturbation, gives
\begin{align} \label{eq:B2}
	\hat{T} &= \hat{U} - i \epsilon \sum_{k=1}^{N} \hat{U}_{N,k+1} \hat{W}_k \hat{U}_{k,1} + \mathcal{O}(t^2) ,
\end{align}
where $\hat{U}_{\ell,k}=\eu{-i\epsilon\hat{V}_{\ell}}\eu{-i\epsilon\hat{V}_{\ell-1}}\cdots\eu{-i\epsilon\hat{V}_{k+1}}\eu{-i\epsilon\hat{V}_k}$ such that $\hat{U}_{N,1}$ is identical to $\hat{U}$ defined in Sec.~\ref{sec:MASHPLDM}.
Note that the second-order correction, which we have neglected, includes $N^2$ terms of order $\mathcal{O}(\epsilon^2)$, such that there is an overall error of $\mathcal{O}(t^2)$.

Next, we introduce mapping variables using the reciprocal properties of any valid inverse kernel:
\begin{align} \label{eq:B3}
	\hat{T} &= \hat{U} - i \epsilon \sum_{k} \int \rmd \S_k \, \hat{U}_{N,k+1} W_k(\S_k) \winv_{\Bq_k}(\S_k) \hat{U}_{k,1} + \mathcal{O}(t^2) ,
\end{align}
where $W_k(\S_k) = \tr[\hat{W}_k\hat{w}_{\Bq_k}(\S_k)]$.
The order in which kernels and propagators appear can be interchanged using a rule derived by rearranging Eq.~\eqref{eq:winv_prop}:
\begin{align} \label{eq:reorder}
    \winv_{\Bq_{k}}\bk[big]{\S(k\epsilon)} \, \eu{-i\epsilon\hat V_{k}}
    &= \eu{-i\epsilon\hat V_{k}} \, \winv_{\Bq_{k-1}}\bk[big]{\S((k-1)\epsilon)} ,
\end{align}
which is valid for any linear inverse kernel such as \eqn{eq:MASH_PLDM_inverse} or the Stratonovich--Weyl kernel.
Using \eqn{eq:reorder} recursively results in the expression
\begin{align} \label{eq:B5}
	\hat{T}
	&= \hat{U} - i \epsilon \sum_{k} \int\rmd \S_k\, \hat{U}_{N,k+1} W_k(\S_k) \hat{U}_{k,1} \winv_{\Bq_0}(\S_k(-k\epsilon)) + \mathcal{O}(t^2) .
\end{align}

We insert the identity $\Id=\int\d\S\,\winv_{\Bq_0}(\S)$ in the first term
and change variables in the second from $\S_k$ to $\S(k\epsilon)$, 
using the fact that rotations are orthogonal transforms, such that $\rmd\S_k = \rmd\S(k\epsilon) = \rmd\S$.
This gives
\begin{align} \label{eq:B6}
	\hat{T}
	&= \int \rmd \S \left[ \hat{U} - i \epsilon \sum_{k} \hat{U}_{N,k+1} W_k(\S(k\epsilon)) \hat{U}_{k,1} \right] \winv_{\Bq_0}(\S) + \mathcal{O}(t^2) .
\end{align}

Finally, we can resum the perturbation series as
\begin{align} \label{eq:T_MASHPLDM}
	\hat{T} &= \int \rmd \S \, \eu{-i\epsilon W_{N}(\S(N\epsilon))} \, \eu{-i\epsilon\hat{V}_{N}}
    \cdots \eu{-i\epsilon W_1(\S(\epsilon))} \, \eu{-i\epsilon\hat{V}_1} \, \winv_{\Bq_0}(\S) + \mathcal{O}(t^2) .
\end{align}
The PLDM approach thus captures the QCLE dynamics exactly to first-order in time, but may introduce an error at longer times.
Nonetheless, it is clear that the resummation captures the scalar part of $\hat{W}_k$ exactly
and so any error introduced must be related to the treatment of the back action of the electronic dynamics on the nuclei.

In fact, for the simple case in which the two adiabatic potentials are uncoupled,
such that $\nac(\q)=\bm{0}$,
the resummation [\eqn{eq:T_MASHPLDM}] is exact to all orders when using MASH-PLDM kernels.
This is easily shown using the fact that the operators in Eq.~\eqref{eq:Tagain} all commute and because $\hat \sigma_z^2=\Id$,
such that all orders of perturbation theory are proportional to $\hat{U}$ or $\hat{U}\hat{\sigma}_z$.
Likewise, using the fact that $\sgn(S_z)^2=1$ (except for a case of zero measure), the MASH-PLDM version of Eq.~\eqref{eq:T_MASHPLDM} can be written in terms of $\hat{U}\int\d\S\,\winv_{\Bq_0}(\S)$ and $\hat{U}\int\d\S\sgn(S_z)\winv_{\Bq_0}(\S)$, both of which evaluate to the correct QCLE result.
Note that the same is not true of spin-PLDM, where $\hat{\sigma}_z$ is represented by $S_z$, for which $S_z^2\ne1$ in general.

Note a subtle mathematical detail that arises in the case of MASH-PLDM\@.
Because $W_k(\S)$ includes a $\delta(S_z)$ term in the presence of nonadiabatic coupling, 
the exponential $\eu{-i\epsilon W_k(\S)}$ is formally not well defined.
Nonetheless, this does not appear to invalidate the following steps of the derivation,
as we provide alternative proofs of the connection between MASH-PLDM and QCLE in Appendix~\ref{app:QCLE}\@.
In fact, we show there that the overall effect of exponentiating the delta function appears to be equivalent to applying the momentum-jump approximation introduced by Kapral and coworkers.\cite{Kapral1999QCLE,Kapral2015QCL}

The derivation we have presented here is quite different from the justifications used in previous PLDM literature.
In particular, Hsieh and Kapral introduced $N$ sets of mapping variables in the forward propagator (and another $N$ sets in the backward propagator) and then employed an approximation that coherent states are orthogonal in order to join them together into a continuous trajectory.\cite{Hsieh2012FBTS}
Unfortunately, as they point out in their paper, coherent states are overcomplete and therefore not orthogonal.
For lack of a better argument, the same approach was used to derive spin-PLDM\@.\cite{spinPLDM1}
In the current work, we obtain a continuous-trajectory picture after introducing only one set of mapping variables in the forward propagator (and another for the backward propagator).
Our new derivation does not invoke the unjustified orthogonalization approximation
and instead relies on perturbation theory.
This explains why PLDM is accurate at short times or for systems with weak nonadiabatic coupling.
Although we have no mathematical analysis on the properties of the method for long times in systems with strong nonadiabatic coupling,
the numerical results presented in Sec.~\ref{sec:results} seem to imply that the method still behaves well, at least in the cases tested.

\section{Justification of the alternative inverse kernels}
\label{app:variant}

A key step in our derivation of PLDM requires \eqn{eq:winv_prop} to hold in order that we can pull the inverse kernel through the propagator in \eqn{eq:B5}.
This is clearly true for the linear version \eqn{eq:MASH_PLDM_inverse}, but not for those with initial or final projections \eqn{eq:variants}.
Nonetheless, we will show that these kernels obey a more general and sufficient relation.

In particular, we will prove that
\begin{align} \label{QED}
    \hat{U}^\dagger\hat{W}\hat{U} =  \int \rmd\S \, W(\mathsf{R}^\T\S) \, \winv(\S),
\end{align}
for any of the proposed inverse kernels,
where $\mathsf{R}$ is a rotation corresponding to the unitary transform $\hat{U}$, and $W(\S) = \tr\sbk{\hat w (\S) \hat W}$ is the mapped function of an arbitrary operator $\hat W$.
Equation~\eqref{QED} is obviously true for the identity component of $\hat{W}$, for which $W$ is independent of $\S$.  We therefore focus on the traceless part of $\hat{W}$ in the following.
For notational convenience, we drop the subscript $\q$ in this section.
It is clear that \eqn{QED} is sufficient to obtain \eqn{eq:B6} from \eqn{eq:B2} by substituting $\hat{U}=\hat{U}_{k,1}$ and inserting $\hat{U}\hat{U}^\dagger=\Id$.

First we prove the relation for the simpler linear case of $\winv_{xyz}(\S)=S_x\hat \sigma_x+S_y\hat \sigma_y+S_z\hat \sigma_z$.  Note that we do not need to consider the identity component if we assume $\hat{W}$ to be traceless, as this will integrate to zero.
This operator is therefore not a valid inverse kernel on its own, but is used to construct the inverse kernels with initial and final projections.
Nonetheless, the linear case is particularly simple, as a unitary operation on $\winv_{xyz}(\S)$ is equivalent to a rotation of the spin vector,
i.e., $\hat{U}^\dagger\winv_{xyz}(\S)\hat{U} = \winv_{xyz}(\mathsf{R}\S)$.\cite{MahlerBook} %
Using this, the proof can be stated as follows:
\begin{subequations} \label{linear_proof}  
\begin{align}
    \hat{U}^\dagger\hat{W}\hat{U}
    &= \hat{U}^\dagger \left[ \int \rmd\S \, W(\S) \winv_{xyz}(\S) \right] \hat{U}
    \\ &= \int \rmd\S \, W(\S) \, \winv_{xyz}(\mathsf{R}\S),
    \\ &= \int \rmd\S_\mathsf{R} \, W(\mathsf{R}^\T\S_\mathsf{R}) \, \winv_{xyz}(\S_\mathsf{R}),
\end{align}
\end{subequations}
where $\S_\mathsf{R} = \mathsf{R}\S$ and we have used the fact that $\mathsf{R}$ is an orthogonal transform and thus has no Jacobian.
Renaming the dummy variable thus proves \eqn{QED} for the linear case.

In order to prove \eqn{QED} for the more difficult (nonlinear) inverse kernel $\winv_\mathrm{init}(\S)$ as defined in \eqn{eq:initial_kernel}, we will make use of the symmetry properties of the integrals.
In particular, note that the integrand in \eqn{QED} is symmetric with respect to $\S\mapsto-\S$.  This relies on the fact that the inversion operator commutes with the rotation, i.e., $\mathsf{R}^\T(-\S)=-\mathsf{R}^\T\S$,
and that $W(\S)$ and $\winv(\S)$ are both antisymmetric.
It is therefore clear that the integrals over the northern and southern hemispheres are equal to each other and, thus, each contributes half of the total integral.
This is sufficient to prove \eqn{QED} for the nonlinear case, as each element of the matrix $\winv_\mathrm{init}(\S)$ is equal to twice that of $\winv_{xyz}(\S)$ in a particular hemisphere and zero in the other.
In particular, $\winv_\mathrm{init}(\S) = 2 \, \winv_{xyz}(\S) \hat{h}(\S)$,
with $\hat{h}(\S) = \ketbra{\Phi_\mathrm{g}}{\Phi_\mathrm{g}}h(-S_z) + \ketbra{\Phi_\mathrm{e}}{\Phi_\mathrm{e}}h(S_z)$.
By denoting the spherical integrals restricted to the lower and upper hemispheres by $\int_\mathrm{g}\d\S$ and $\int_\mathrm{e} \d\S$,
we find
\begin{subequations}
    \begin{align}
        \int \d \S \, W(\mathsf{R}^\T\S) \, \winv_\mathrm{init}(\S)
        &= 2 \int \d \S \, W(\mathsf{R}^\T\S) \, \winv_{xyz}(\S) \hat{h}(\S) \\
        &= 2 \int_{\mathrm{g}} \d \S \, W(\mathsf{R}^\T\S) \, \winv_{xyz}(\S) \ketbra{\Phi_\mathrm{g}}{\Phi_\mathrm{g}} + 2 \int_{\mathrm{e}} \d \S \, W(\mathsf{R}^\T\S) \, \winv_{xyz}(\S) \ketbra{\Phi_\mathrm{e}}{\Phi_\mathrm{e}} \\
        &= \int \d \S \, W(\mathsf{R}^\T\S) \, \winv_{xyz}(\S) \ketbra{\Phi_\mathrm{g}}{\Phi_\mathrm{g}} + \int \d \S \, W(\mathsf{R}^\T\S) \, \winv_{xyz}(\S) \ketbra{\Phi_\mathrm{e}}{\Phi_\mathrm{e}}\\
        &= \int \d \S \, W(\mathsf{R}^\T\S) \, \winv_{xyz}(\S),
    \end{align}
\end{subequations}
which thus proves \eqn{QED} in the general case.
Analogously, this can be shown for the alternative variant of the inverse kernel in \eqn{eq:final_kernel}, $\winv_\mathrm{fin}(\S) = 2 \hat{h}(\S) \winv_{xyz}(\S)$.

In this way, the MASH-PLDM method with projections at the initial or final time can be written as
\begin{subequations}
\begin{align}
    C_{AB}^\text{init}(t) &= 4 \int \frac{\d \Bq \d \Bp}{(2\pi)^f} \d\S \d\S' \, \rho_\text{n}(\Bq, \Bp) \tr \sbk{\hat A \hat{h}_{\Bq}(\S') \winv_{\Bq,{xyz}}(\S') \hat{U}^\dagger \hat B \hat{U} \winv_{\Bq,{xyz}}(\S) \hat{h}_{\Bq}(\S)},
    \label{eq:MASH-PLDM_initial}
    \\
    C_{AB}^\text{fin}(t) &= 4 \int \frac{\d \Bq \d \Bp}{(2\pi)^f} \d\S \d\S' \, \rho_\text{n}(\Bq, \Bp) \tr \sbk{\hat A  \winv_{\Bq,{xyz}}(\S') \hat{U}^\dagger \hat{h}_{\Bq(t)}(\S'(t)) \hat B \hat{h}_{\Bq(t)}(\S(t)) \hat{U} \winv_{\Bq,{xyz}}(\S)},
    \label{eq:MASH-PLDM_final}
\end{align}
\end{subequations}
where in the latter, we used \eqn{eq:reorder} to bring the linear part of the kernel back to the start.

\section{Connection between PLDM and QCLE}
\label{app:QCLE}

Here, we investigate the connection between various forms of PLDM and QCLE in the short-time limit.
It is clearly true that the correlation functions are equal at $t=0$, so we focus on the first time derivative, for which QCLE gives
\begin{align} 
    \dot{C}_{AB}^\text{QCLE} &= \int \frac{\d \Bq \, \d \Bp}{(2\pi)^f} \, \rho_\text{n}(\Bq, \Bp) \tr \sbk{\hat A \, \der{\hat{B}}{t}} .
    \label{eq:QCLECdot}
\end{align}
Under the QCLE equations of motion,\cite{Kapral1999QCLE} the time-derivative of a partially Wigner-transformed operator
$\hat{B}=B_\mu(\Bq,\Bp) \pauli_\mu(\Bq)$, where $\pauli_\mu$ is any Pauli matrix or the identity, is
\begin{subequations} \label{eq:QCLEdBdt}
\begin{align} 
    \der{\hat{B}}{t} &= i\left[\hat{V},\hat{B}\right] + \pder{\hat{B}}{\Bq}\cdot\frac{\Bp}{m} + \half \left[\pder{\hat{B}}{\Bp}, \hat{\bm{F}}(\Bq)\right]_+
    \\ &= i\left[\hat{V} + \frac{\nac(\Bq)\cdot\Bp}{m}\pauli_y, \hat{B}\right] + \pauli_\mu \pder{B_\mu}{\Bq}\cdot\frac{\Bp}{m} 
    + \half \pder{B_\mu}{\Bp} \cdot \left[\pauli_\mu, \hat{\bm{F}}(\Bq)\right]_+ ,
\end{align}
\end{subequations}
where the force operator is defined as $\hat{\bm{F}}(\Bq)=-\nabla\hat{V}(\Bq)$ [see \eqn{eq:force}]
and we have used the spatial derivative of the Pauli operator given below \eqn{eq:pauli}.
Note that, if $\hat{B}$ contains a coherence operator (e.g., $\pauli_\mu=\pauli_x$), the forces will be averaged over the two states in the anticommutator $[\cdot,\cdot]_+$.
This justifies the use of trajectories following the averaged PES in the MASH-PLDM method.

The equivalent object within the PLDM framework (with any valid choice of mapping kernels) is defined as
\begin{align} \label{eq:PLDM_ddt}
    \dot{C}_{AB}^\text{PLDM} &= \int \frac{\d \Bq \, \d \Bp}{(2\pi)^f} \, \d\S \, \d\S' \, \rho_\text{n}(\Bq, \Bp) \tr \sbk{\hat A  \winv_{\Bq}(\S') \der{\hat{U}^\dagger\hat{B}\hat{U}}{t} \winv_{\Bq}(\S)}_{t=0} ,
\end{align}
where under the PLDM equations of motion,
\begin{subequations}
\begin{align}
    \left.\der{\hat{U}^\dagger\hat{B}\hat{U}}{t}\right|_{t=0} &= \left[\dot{\hat{U}}^\dagger \hat{B} \hat{U} + \hat{U}^\dagger \hat{B} \dot{\hat{U}} + \hat{U}^\dagger \dot{\hat{B}}\hat{U}\right]_{t=0}
    \\ &= i\left[\hat{V},\hat{B}\right] + \pder{\hat{B}}{\Bq} \cdot \dot{\Bq} + \pder{\hat{B}}{\Bp} \cdot \dot{\Bp}
    \\ &= i\left[\hat{V},\hat{B}\right] + \pder{\hat{B}}{\Bq} \cdot \frac{\Bp}{m} + \pder{\hat{B}}{\Bp} \cdot \bm{F}(\Bq,\S,\S')
    \\ &= i\left[\hat{V} + \frac{\nac(\Bq)\cdot\Bp}{m}\pauli_y,\hat{B}\right] + \pauli_\mu \pder{B_\mu}{\Bq} \cdot \frac{\Bp}{m} +  \pauli_\mu \pder{B_\mu}{\Bp} \cdot \bm{F}(\Bq,\S,\S').
\end{align}
\end{subequations}
Here, we used $\hat{U}=\Id$ and $\dot{\hat{U}}=-i\hat{V}$ in the $t\rightarrow0$ limit.
The first two terms clearly recover the correct QCLE dynamics as the inverse kernels in Eq.~\eqref{eq:PLDM_ddt} integrate to the identity.
The third term involving the force can also be shown to match with the QCLE result
using the fact that 
$\bm{F}(\Bq,\S,\S') = \half ( \tr[\w_{\Bq}(\S) \hat{\bm{F}}] + \tr[\w_{\Bq}(\S') \hat{\bm{F}}])$
along with the property $\int \d\S \tr[\w_{\Bq}(\S) \hat{\bm{F}}] \winv_{\Bq}(\S) = \hat{\bm{F}}$. %
Because in \eqn{eq:PLDM_ddt}, one inverse kernel is on the right-hand-side of the $\hat{B}$ operator and the other on the left, this reproduces the anticommutator of Eq.~\eqref{eq:QCLEdBdt}.
Overall, this proves that PLDM captures the correct short-time dynamics (up to first-order in time) for any consistent choice of mapping kernels, such as for
spin-PLDM.\cite{spinPLDM1,spinPLDM2} %

Note that this proof does not apply to the original PLDM of Huo and Coker,\cite{Huo2011densitymatrix,Huo2012PLDM}
which (as pointed out by Hsieh and Kapral)\cite{Hsieh2012FBTS} did not separate the trace component of the potential operator before mapping.
In particular, 
its kernel (which is based on Meyer--Miller--Stock--Thoss mapping) does not obey the normalization condition $\tr[\hat{w}(\S)]=1$,\cite{identity} %
where $\S$ represents the appropriate set of mapping variables.
This means that
the identity component of the force term depends in a complicated way on the mapping variables,
and that when multiplied by the inverse kernel and integrated, it does not recover the QCLE force.
This problem was circumvented in FBTS by removing the trace component of the force and treating it as a scalar along with the kinetic energy,\cite{Hsieh2012FBTS}
which thus also reduces to QCLE in the short-time limit.

Importantly, this proof is not valid for the case of MASH-PLDM, %
as although the correlation function is differentiable, the individual trajectories are not (due to the hops) and thus, one cannot reverse the order of differentiation and integration to obtain \eqn{eq:PLDM_ddt}.
For this reason, we now consider a different form of the QCLE which includes momentum jumps.

\subsection{QCLE under the momentum-jump approximation}

The QCLE equations of motion can be alternatively written as
\begin{align}
    \der{\hat{B}}{t}
    &=
    i\left[\hat{V}, \hat{B}\right]
    + \pauli_\mu \pder{B_\mu}{\Bq}\cdot\frac{\Bp}{m} 
    - \left(\pauli_\mu \nabla V_0 + \thalf \left[\pauli_\mu,\pauli_z\right]_+ \nabla V_z \right) \cdot \pder{B_\mu}{\Bp} + \hat{\mathcal{J}}_\mu B_\mu,
\end{align}
with the coupling operator\cite{Kapral1999QCLE,Kapral2015QCL}
\begin{align}
    \hat{\mathcal{J}}_\mu
    &= i\left[\pauli_y, \pauli_\mu\right] \frac{\nac\cdot\Bp}{m} + \thalf \left[\pauli_\mu,\pauli_x\right]_+ 4 V_z \nac\cdot\Bp \pder{}{\mathcal{Y}} ,
\end{align}
where $\mathcal{Y}=(\Bp\cdot\nac(\Bq)/||\nac(\Bq)||)^2$, which is proportional to available kinetic energy.
Under the momentum-jump approximation (MJA),\cite{Kapral1999QCLE,Kapral2015QCL,Kapral2016FSSH} this operator becomes
\begin{align}
    \hat{\mathcal{J}}_\mu^\text{MJA} &= i\left[\pauli_y, \pauli_\mu\right] \frac{\nac\cdot\Bp}{m} \frac{\hat{\jmath}_+ + \hat{\jmath}_-}{2} + \left[\pauli_\mu,\pauli_x\right]_+ \frac{\nac \cdot \Bp}{m} \frac{\hat{\jmath}_+ - \hat{\jmath}_-}{2} ,
\end{align}
where the momentum-jump operators are
\begin{align}
    \hat{\jmath}_\pm = \exp\left(\pm 2 m V_z \pder{}{\mathcal{Y}}\right) .
\end{align}
Note that $\hat{\jmath}_\pm \mathcal{Y} = \mathcal{Y} \pm 2mV_z$ are generators of translations of $\mathcal{Y}$
and therefore
$\hat{\jmath}_\pm B_\mu(\Bp_\text{old}) = B_\mu(\Bp_\text{new}^\pm)$ has the effect of performing a momentum rescaling for a hop up or a hop down [see \eqn{momentum_rescaling}].

For an important example, consider the short-time limit of the correlation function with $\hat{A}=\pauli_x$ and $\hat{B}=B_0(\p)\otimes\Id$,
which gives
\begin{align} \label{QCLEMJA}
    \dot{C}_{AB}^\text{QCLE-MJA}
    &= 2 \int \frac{\d \Bq \, \d \Bp}{(2\pi)^f} \, \frac{\nac\cdot\Bp}{m} \left[B_0(\Bp_\text{new}^+) - B_0(\Bp_\text{new}^-)\right] .
\end{align}

\subsection{Short-time behavior of MASH-PLDM}

We now study the MASH-PLDM correlation function with $\hat{A}=\pauli_x$ and $\hat{B}=B_0(\p)\otimes\Id$ in the short-time limit.
We first consider the variant with the linear inverse kernel, where evaluating the trace gives
\begin{align}
    \tr\left[\pauli_x\winv_{\Bq,\text{lin}}(\S')\hat{U}^\dagger\Id\hat{U}\winv_{\Bq,\text{lin}}(\S)\right]_{t=0}
    &= S_x + S_x' + 2i(S_z S_y' - S_y S_z') .
\end{align}
Through the symmetry of exchanging the spins,
the imaginary contributions cancel %
and leave the integral $\int \d\S\,\d\S'\, 2 S_x \, B_0(\Bp(t))$ for each nuclear initial condition.
The correlation between $S_x$ and $\Bp(t)$
picks out trajectories which hop due to the $\S$ spin changing hemisphere, whereas the integral over $\S'$ simply gives a factor of 2.
We assume that the nuclei have sufficient energy that the hops are not frustrated.

To first-order in time, $S_z(t) \simeq S_z - \frac{2\nac \cdot \Bp}{m} S_x t$,
so that integrating over $S_z$ (using $\int_{-1}^1\d S_z\cdots=\int_0^\pi\d\theta\sin\theta\cdots$) gives a contribution from trajectories which hop down of $R(\frac{2\nac\cdot\Bp}{m}S_x)t$, where $R(x)=\max(0,x)$ %
is the ramp function, and similarly for trajectories which hop up.
Therefore,
\begin{align}
    C_{AB}^\text{lin}(t)
    &= 4t \int \frac{\d\Bq \, \d\Bp}{(2\pi)^f} \, \frac{\d\varphi}{2\pi} \, \rho_\text{n}(\Bq, \Bp) \, S_x \left[R\left(\frac{2\nac\cdot\Bp}{m}S_x\right) B_0(\Bp_\text{new}^+) - R\left(-\frac{2\nac\cdot\Bp}{m}S_x\right) B_0(\Bp_\text{new}^-)\right] + \mathcal{O}(t^2)
    \\
	 &= 2t \int \frac{\d \Bq \, \d \Bp}{(2\pi)^f} \, \rho_\text{n}(\Bq, \Bp) \, \frac{\nac\cdot\Bp}{m} \left[ B_0(\Bp_\text{new}^+) - B_0(\Bp_\text{new}^-)\right] + \mathcal{O}(t^2) ,
\end{align}
which is in agreement
with the QCLE-MJA result given by \eqn{QCLEMJA}.

We proceed with the variant of MASH-PLDM using the initial inverse kernel.
Evaluating the trace and making use of the exchange symmetry of the spins we find
\begin{align}
    C_{AB}^\text{init}(t) = 8 \int \frac{\d \Bq \, \d \Bp}{(2\pi)^f} \, \d\S \, \d\S' \, \rho_\text{n}(\Bq, \Bp) \int_\mathrm{e}\d\S\int_\mathrm{g} \d\S' \bk{S_z S_x' - S_x S_z'} B_0(\Bp(t)) ,
\end{align}
where $\int_\mathrm{g/e}\d\S \cdots$ denote the spherical integrals restricted to the lower and upper hemispheres, respectively.
Performing the integral over one spin $\int_\mathrm{e/g} \d \S\, S_z = \pm\frac12$ puts the remaining integral into the same form as that for the linear kernel.
Therefore, we come to the same conclusion for the short-time behavior of the initial inverse kernel MASH-PLDM.

For the variant using the final inverse kernel,
we use $\hat{U}\winv_{\Bq,xyz}(\S)=\winv_{\Bq(t),xyz}(\S(t))\hat{U}$ to find
\begin{align}
    C_{AB}^\text{fin}(t) = 8 \int \frac{\d \Bq \, \d \Bp}{(2\pi)^f} \, \rho_\text{n}(\Bq, \Bp) \Big[\int_\mathrm{e}\d\S(t)\int_\mathrm{e} \d\S'(t) - \int_\mathrm{g}\d\S(t)\int_\mathrm{g} \d\S'(t) \Big] S_x(t) S_z'(t) B_0(\Bp(t)) + \mathcal{O}(t^2),
\end{align}
where we have %
substituted variables to rewrite the integrals in terms of the final spin vectors, and neglected the time dependence of the evolution operator, which cannot contribute to this correlation function to first-order in time.
Evaluating the integrals as before results in the same conclusion as for the other two variants of MASH-PLDM.

We have therefore demonstrated that all three versions of MASH-PLDM recover the correct short-time limit of the QCLE under the momentum-jump approximation.

\section{Linearized MASH}
\label{app:MASHLSC}

Another property of the mapping procedure is that one can also map an operator using the inverse kernel (known as a dual) to give $A_{\q}^\text{dual}(\S) = \tr\sbk{\hat A \, \winv_{\q}(\S)}$.\cite{Stratonovich1957,spinmap}
It is then possible to show that $\tr[\hat A \hat B] = \int \d\S \, A_{\q}^\text{dual}(\S) B_{\q}(\S)$ for any operators $\hat A$ and $\hat B$ in the electronic Hilbert space.

For this purpose, we propose the inverse mapping kernel
\begin{equation} \label{eq:dual}
    \hat w_{\q,\text{dual}}^{-1}(\S) = \frac12\bk{2|S_z|\Id + 2 S_x \hat \sigma_x(\q) + 2 S_y \hat \sigma_y(\q) + 2 S_z \hat \sigma_z(\q)},
\end{equation}
for which the dual mappings of Pauli matrices are simply $\sigma_\mu^\mathrm{dual}(\S) = 2S_\mu$ and the dual mapping of the identity is $2|S_z|$.
It is easy to check that these functions obey the properties given above.
For example, if $\hat{A}=\hat{B}=\hat\sigma_z$, then $\tr[\hat\sigma_z\hat\sigma_z]=\int \d\S\, 2 S_z \sgn(S_z) = 2$, which is the correct result.

Following the formalism of LSCIVR\cite{Miller2009mapping} and spin-LSC methods,\cite{spinmap}
we can construct a fully linearized variant of MASH-PLDM (which we call MASH-LSC).
This method uses only one mapping variable, $\S$, and instead of mapping the propagators, maps the operator $\hat B$ with the mapping kernel \eqn{eq:MASH_PLDM_kernel} and $\hat A$ with the inverse (dual) kernel \eqn{eq:dual}:
\begin{align}
    C_{AB}^\text{MASH-LSC}(t) &= \int \frac{\d \Bq \, \d \Bp}{(2\pi)^f} \, \d \S \, \rho_\text{n}(\q, \p) \, A_{\q}^\text{dual}(\S) B(\q(t),\p(t),\S(t)).
\end{align}
Note that we find that the good properties highlighted below hold for this choice, but not if we choose to map $\hat{B}$ with the dual instead of $\hat{A}$.

This MASH-LSC method uses the force $\bm{F}(\q,\S)$ [\eqn{eq:MASHforce}] exactly as in the original MASH approach.
In fact, the only difference between MASH-LSC and the original MASH is in the way that correlation functions are constructed.
The original MASH approach was written in a slightly different form using weighting functions, $\mathcal{W}_{AB}$.\cite{MASH}
MASH-LSC can also be written in this form
with the alternative set of weighting functions
$\mathcal{W}_\mathrm{PP}=2|S_z|$,
$\mathcal{W}_\mathrm{PC}=4|S_z|\delta(S_z(t))$,
$\mathcal{W}_\mathrm{CP}=2$, and
$\mathcal{W}_\mathrm{CC}=4\delta(S_z(t))$,
where $\mathrm{P}\in\{\Id,\hat{\sigma}_z\}$ and $\mathrm{C}\in\{\hat{\sigma}_x,\hat{\sigma}_y\}$.
However, although $\mathcal{W}_\mathrm{PP}$ and $\mathcal{W}_\mathrm{CP}$ are identical to those of the original MASH method, $\mathcal{W}_\mathrm{PC}$ and $\mathcal{W}_\mathrm{CC}$ are different.
In particular, correlation functions where $\hat{B}$ is a coherence only have contributions from trajectories whose spin vectors end exactly on the equator.
This means that only a few trajectories will contribute at any time, making the statistical convergence slower, unless tricks can be devised to enhance the sampling, such as by running the trajectories backwards in time,\cite{MASHEOM,MASHTPS,MASHTPSDellago} having started them at the equator.
Nonetheless, despite the practicality issues, we can study the mathematical properties of the method.

In particular, it is interesting to investigate whether these alternative weighting functions obey the same properties as the original MASH (and may even lead to improved accuracy), as it is known that the MASH weighting functions are not unique.\cite{Vavrin,MASHRedfield}
When $\hat{B}$ is a population operator, the MASH-LSC case is identical to ordinary MASH and thus recovers QCLE to first-order in time.\cite{MASH,MASHreview}
It is possible to show that this good behavior also holds for coherence--coherence correlation functions, such as with $\hat{A}=\pauli_x$ and $\hat{B}=\pauli_y$, for which 
after linearizing the time, we obtain for each initial nuclear condition
\begin{subequations}
\begin{align}
    \int \d \S \, A_{\q}^\text{dual}(\S) B(\q(t),\p(t),\S(t))
    &= \int \d \S \, 2 S_x 2\delta(S_z(t)) S_y(t)
    \\ &\simeq \int \d \S \, 2 S_x 2\delta\left(S_z - 2\frac{\nac\cdot\p}{m}S_xt\right) \left[S_y + 2V_z S_x t\right]
    \\ &= 4 V_z t + \mathcal{O}(t^2) ,
\end{align}
\end{subequations}
which is in agreement with QCLE to first order in time\@.
In fact, all the coherence--coherence correlation functions can be shown to give the correct short-time limits for any valid choice of kernel and its dual.

We next consider the case of $\hat{A}=\pauli_z$ and $\hat{B}=B_x\otimes\pauli_x$, for which
\begin{subequations}
\begin{align}
    \int \d \S \, A_{\q}^\text{dual}(\S) B(\q(t),\p(t),\S(t))
    &= \int \d \S \, 2S_z 2\delta(S_z(t)) S_x(t) B_x(\q(t),\p(t))
    \\ &\simeq \int \d \S \, 2S_z 2\delta\left(S_z - 2\frac{\nac\cdot\p}{m} S_x t\right) S_x B_x(\q,\p)
    \\ &= \int \d \S \, 2\left(2\frac{\nac\cdot\p}{m}S_xt\right) 2\delta\left(S_z - 2\frac{\nac\cdot\p}{m} S_x t\right) S_x B_x(\q,\p)
    \\ &= 4 \frac{\nac\cdot\p}{m} B_x(\q,\p) t + \mathcal{O}(t^2) ,
\end{align}
\end{subequations}
which leads to the correct QCLE result in the short-time limit.

Finally, the case of $\hat{A}=\Id$ and $\hat{B}=B_x\otimes\pauli_x$ is the most interesting as correlations only appear for trajectories which end in a hop, which rescales the momentum in the direction of the nonadiabatic coupling vector as $\tilde{p}_\text{new} = \sqrt{(\tilde{p}_\text{old})^2 + 2mV_z\sgn(S_x)}$.
Here we also make the assumption that $(\tilde{p}_\text{old})^2 \gg 2mV_z$, such that 
\begin{subequations}
\begin{align}
    \int \d \S \, A_{\q}^\text{dual}(\S) B(\q(t),\p(t),\S(t))
    &= \int \d \S  \, 2|S_z| 2\delta(S_z(t)) S_x(t) B_x(\q(t),\p(t))
    \\ &\simeq \int \d \S \, 2|S_z| 2\delta(S_z(t)) S_x \pder{B_x}{\p}\cdot[\p(t) - \p]
    \\ &\simeq \int \d \S \, 2\left|2\frac{\nac\cdot\p}{m}S_xt\right| 2\delta(S_z(t)) S_x \, \pder{B_x}{\p}\cdot\nac \, \frac{mV_z\sgn(S_x)}{|\nac\cdot\p|}
    \\ &\simeq 4 V_z t \, \nac\cdot\pder{B_x}{\p} + \mathcal{O}(t^2) ,
\end{align}
\end{subequations}
which is the same as QCLE in the limit that $2mV_z\ll \tilde{p}^2$.

Other population--coherence correlation functions are zero according to both QCLE and MASH-PLDM\@.
We have therefore derived a new version of MASH that %
reduces to the QCLE in the short-time limit as long as the kinetic energy is much larger than the energy gap.
Although this point was not fully appreciated in the first publications,\cite{MASH,MASHreview}
the original MASH method also only rigorously captures QCLE in these limits.\cite{piMASH}

\bibliography{references,references_new}

\clearpage
\section*{Supplementary Material}
\renewcommand{\thefigure}{S\arabic{figure}}
\setcounter{figure}{0}
    \section*{Convergence of MASH-PLDM}
\label{app:convergence}
\begin{figure}[h]
    \centering
    \includegraphics[width=0.85\linewidth]{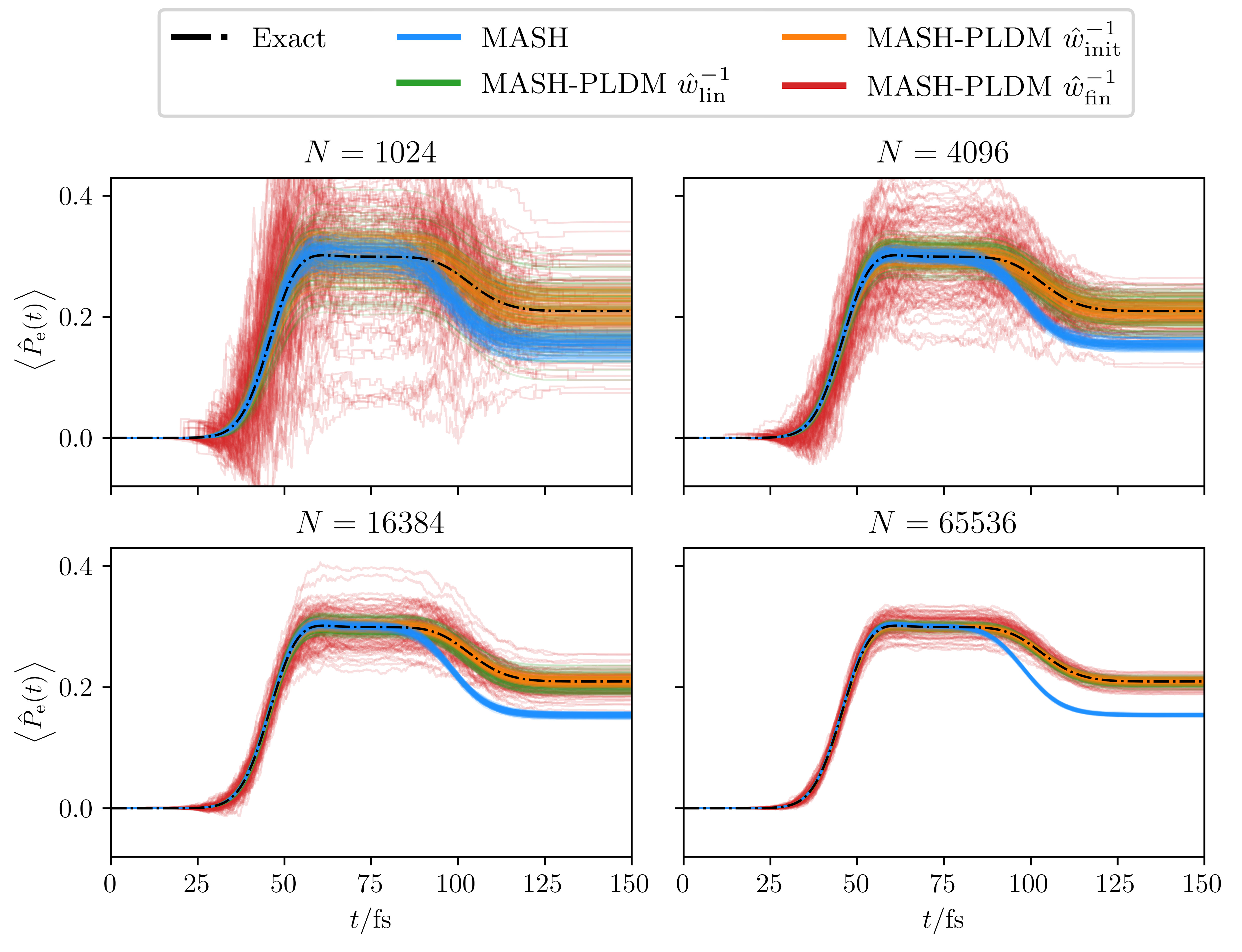}
    \caption{Excited-state population over 100 trials with $N$ trajectories each for the Tully III model. %
    }
    \label{fig:tully3_100_trials}
\end{figure}

In order to assess the convergence properties of MASH-PLDM, we ran 100 independent simulations of the Tully III model as defined in the main text with $N = 2^{8}, 2^{10},\dots, 2^{18}$ trajectories each.
For this analysis, we considered the time-dependent excited-state population observable $\hat{B}=\hat P_\mathrm{e}=\ketbra{\Phi_\mathrm{e}}{\Phi_\mathrm{e}}$ and compared it to original MASH in Fig.~\ref{fig:tully3_100_trials}.
It can be seen that the linear and initial variants of MASH-PLDM converge faster than the final variant (red), but that the convergence is only slightly slower than the original MASH approach.

\begin{figure}
    \centering
    \includegraphics[width=0.5\linewidth]{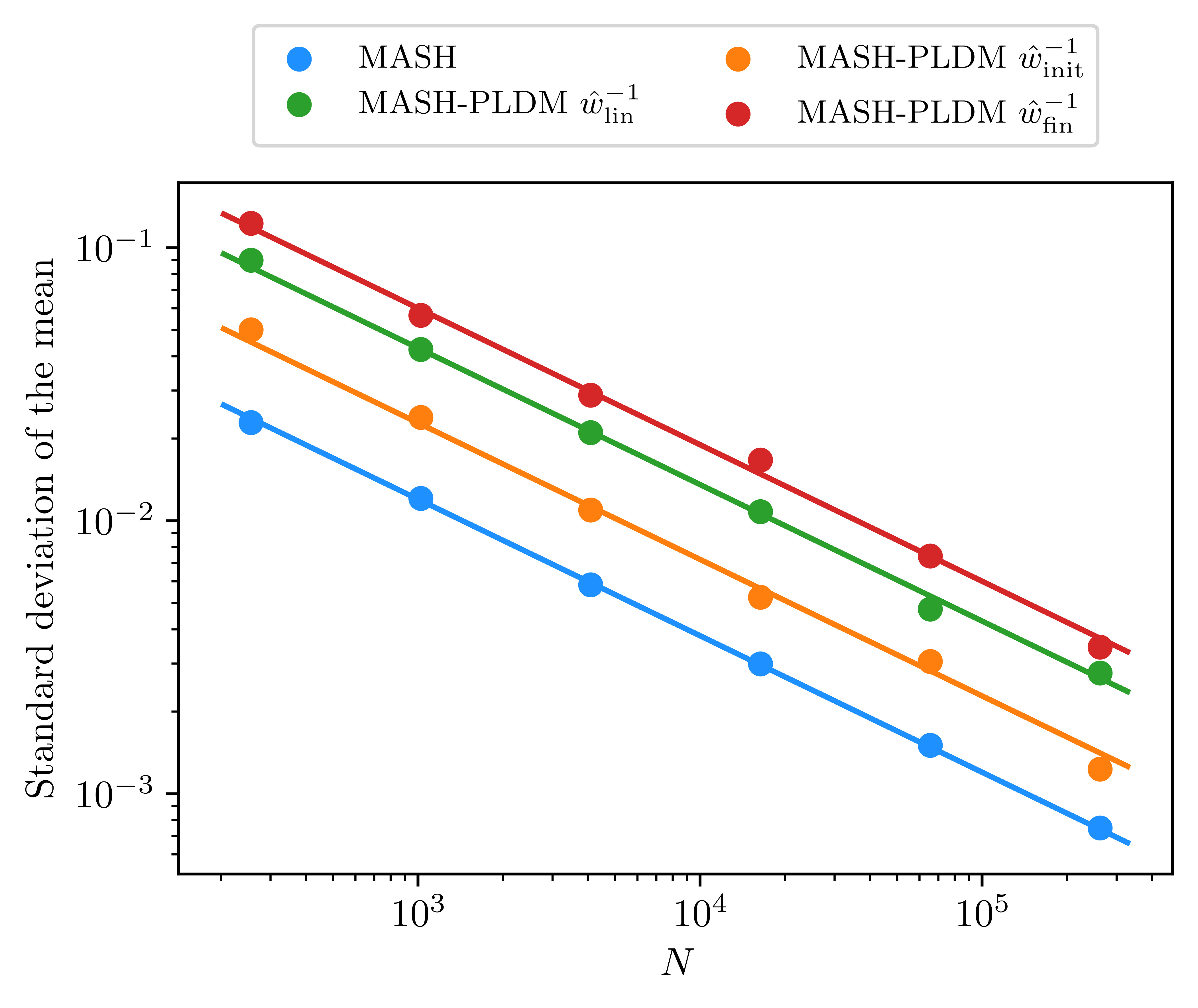}
    \caption{Standard deviation of the mean over 100 trials with $N$ trajectories each for the excited-state population at $t=\SI{150}{fs}$ in the Tully III model.
    Straight lines show $\sigma/\sqrt{N}$ behavior with $\sigma$ chosen to best fit the data points.
    }
    \label{fig:convergence}
\end{figure}

The convergence rate of each method is quantified in Fig.~\ref{fig:convergence}, where it is seen to be well described by a $1/\sqrt{N}$-behavior, as expected for a Monte Carlo approach.
We fitted the standard error of the mean to $\sigma / \sqrt{N}$ and found $\sigma = 0.38, 1.36, 0.72, 1.90$ for MASH and the linear, initial and final variants of MASH-PLDM, respectively.
For a targeted standard error we thus conclude that %
the linear, initial and final variants of MASH-PLDM require 13, 3.6 and 25 times more samples than MASH, respectively.
Note, that this data takes into account the fact that the electronic subsystem is initialized in the ground state, such that both MASH-PLDM with the initial inverse kernel $\winv_\mathrm{init}$ and the original MASH method allow for a restriction of the initial spins to the lower hemispheres, reducing the sampling volume to a quarter and a half, respectively.

\end{document}